\documentclass[%
 reprint,
 amsmath,amssymb,
 aps,
 pra,
]{revtex4-1}

\usepackage{graphicx}
\usepackage{dcolumn}
\usepackage{bm}
\usepackage{hyperref}

\begin{document}

\preprint{APS/123-QED}

\title{Grain extraction and microstructural analysis method for two-dimensional poly and quasicrystalline solids}

\author{Petri Hirvonen}
\email{petri.hirvonen@aalto.fi}
\affiliation{QTF Centre of Excellence, Department of Applied Physics, Aalto University School of Science, P.O. Box 11000, FIN-00076, Aalto, Espoo, Finland}

\author{Gabriel Martine La Boissoni\`ere}
\affiliation{Department of Mathematics and Statistics, McGill  University,  Montreal,  QC  H3A  0B9,  Canada}

\author{Zheyong Fan}
\affiliation{QTF Centre of Excellence, Department of Applied Physics, Aalto University School of Science, P.O. Box 11000, FIN-00076, Aalto, Espoo, Finland}

\author{Cristian-Vasile Achim}
\affiliation{Water Research Center for Agriculture and Mining (CRHIAM), University of Concepci\'on, 4030000 Concepci\'on, Chile}

\author{Nikolas Provatas}
\affiliation{Department of Physics and Centre for the Physics of Materials, McGill University, Montreal, QC H3A 0B9, Canada}

\author{Ken R. Elder}
\affiliation{Department of Physics, Oakland University, Rochester, MI 48309, USA}

\author{Tapio Ala-Nissila}
\affiliation{QTF Centre of Excellence, Department of Applied Physics, Aalto University School of Science, P.O. Box 11000, FIN-00076, Aalto, Espoo, Finland}
\affiliation{Interdisciplinary Centre for Mathematical Modelling and Department of Mathematical Sciences, Loughborough University, Loughborough, Leicestershire LE11 3TU, UK}

\date{\today}

\begin{abstract}
While the microscopic structure of defected solid crystalline materials has significant impact on their physical properties, efficient and accurate determination of a given polycrystalline microstructure remains a challenge. In this paper we present a highly generalizable and reliable variational method to achieve this goal for two-dimensional crystalline and quasicrystalline materials. The method is benchmarked and optimized successfully using a variety of large-scale systems of defected solids, including periodic structures and quasicrystalline symmetries to quantify their microstructural characteristics, \textit{e.g.}, grain size and lattice misorientation distributions. We find that many microstructural properties show universal features independent of the underlying symmetries.
\end{abstract}

\maketitle


\section{Introduction}

The properties of matter in its solid, crystalline state are typically dictated not only by the elemental composition and lattice structure but also the microstructure, i.e. the distribution of grains and lattice defects. The microstructure can have a great influence on mechanical \cite{ref-callister, ref-hull-bacon, ref-gb-strength}, thermal \cite{ref-bimodal, ref-kapitza, ref-thermoelectric}, electrical \cite{ref-electron-transport-graphene, ref-thermoelectric} and other physical properties of the solid phase \cite{ref-graphene-functionalization}. However, mapping the exact relationships between the atomistic details of the microstructure and the more macroscopic material properties is a major challenge -- realistic microstructures are often very complicated and even isolated defects such as grain boundaries or triple junctions have a large number of degrees of freedom to be investigated \cite{ref-hull-bacon, ref-king}. Regardless, realistic model systems and detailed knowledge of the distributions of grains and defects are paramount to this task.

Modeling the formation of realistic microstructures -- a prerequisite to investigate the connections between microstructure and material properties -- is a formidable challenge due to the complex elastic interactions between defects and the vast range of length and time scales involved. While some progress has been made using traditional atomistic modeling methods such as accelerated molecular dynamics \cite{ref-amd}, the recently developed phase-field crystal (PFC) approach is a strong contender. PFC models naturally incorporate diffusion and elastoplasticity in defected crystalline materials and have been shown to produce realistic microstructures for selected materials \cite{ref-backofen, ref-multiscale, ref-gabriel}. Their formulation allows modeling the slow evolution of microstructures with atomic-level resolution in systems of up to mesoscopic size.

Characterizing and analyzing microstructures remains a very difficult task, however. While there exist several methods including variational \cite{ref-var1, ref-var2, ref-var3, ref-var4} and geometric \cite{ref-ovito} to detect the lattice orientation in a polycrystalline material, there have only been few attempts to further extract and measure the network of grains as in Ref. \cite{ref-backofen}. Notably, fully atomistic approaches \cite{ref-detection-3d, ref-gabriel} have been developed to solve both problems by first assigning an orientation to atoms based on their local environment and then assigning them to appropriate grains in an iterative fashion.

Another open issue concerns aperiodic crystalline structures. In particular, the microstructures of quasicrystals and their impact on physical properties are not well known. Quasicrystals are a group of materials that show no long-range translational order but display long-range orientational order, which makes structural analysis a major challenge with traditional means. In particular, they can have, for example, 5-, 8-, 10- or 12-fold rotational symmetries which are not possible in regular periodic crystals. First discovered in 1984, quasicrystals are today known to form a family of hundreds of metallic alloys and soft-matter systems. Quasicrystals have many potential applications due to their low coefficient of friction, resistance to oxidation \cite{ref-qc-oxidation}, and are also attractive in catalytic \cite{ref-qc-catalysis} and epitaxial \cite{ref-qc-epitaxy} applications. Modeling quasicrystals and their evolution using the PFC approach shows great promise. Recent works have considered quasicrystal growth modes \cite{ref-cristian-growth-modes}, interfaces between quasicrystalline grains from multiple separate seeds \cite{ref-cristian-qc-interfaces}, monolayers on quasicrystalline surfaces \cite{ref-monolayer-on-qc} and even three-dimensional quasicrystalline systems \cite{ref-3d-qc}. On the other hand, where periodic crystals display an endlessly repeating motif quasicrystals do not obey this rule which drastically complicates both the detection of a lattice orientation and grain extraction with the current methods \cite{ref-backofen,ref-detection-3d,ref-gabriel}. To our knowledge, no attempts toward grain extraction in quasicrystals have been reported.

In this work, we present and benchmark a powerful variational method for extracting individual grains and analyzing the microstructure in two-dimensional (2D) poly(quasi)crystalline systems from large-scale PFC grain coarsening simulations. We consider both regular square and hexagonal lattice types, as well as quasicrystals with 10- and 12-fold rotational symmetries. We study the sizes, aspect ratios, circularities and neighbor counts of individual grains, as well as the size ratios, misorientations and misalignments between neighboring grains. We demonstrate that the method can be reliably used to quantify the microstructure of 2D crystals and quasicrystals. 

The remainder of this work is organized as follows: Section \ref{sec-methods-grain-extraction} introduces the grain extraction method and Sec. \ref{sec-methods-model-systems} describes the present model systems and the PFC model used to characterize them. In Sec. \ref{sec-assessment}, the performance of the grain extraction method is evaluated and in Sec. \ref{sec-analysis} results of microstructural analysis of different (quasi-)lattice types are given. Section \ref{sec-conclusions} concludes and summarizes our results. Appendix \ref{sec-appendix} gives more details of our methods and additional results.

\section{Methods}
\label{sec-methods}

\subsection{Grain extraction method}
\label{sec-methods-grain-extraction}

The grain extraction method proposed here consists of four steps. In the first step, a density field describing a crystalline or quasicrystalline 2D system is transformed into an "orientation field" indicating the crystallographic orientation and crystalline order at each point. In this work, for the sake of concreteness and ease of implementation we consider mainly PFC generated density fields, but virtually any data containing the spatially distributed atomic density is acceptable; see Sec. \ref{sec-further-applications} in Appendix for examples. Next, a "deformation field" is constructed from the orientation field, highlighting the grain boundaries and isolated dislocations. Then, the system is segmented into "subdomains" via level-fill growth in the deformation field. As the final step, some subdomains need to be merged to recover a structure closer to the true network of grains. This subsection describes these steps in detail.

\begin{figure*}
\includegraphics[width=\textwidth]{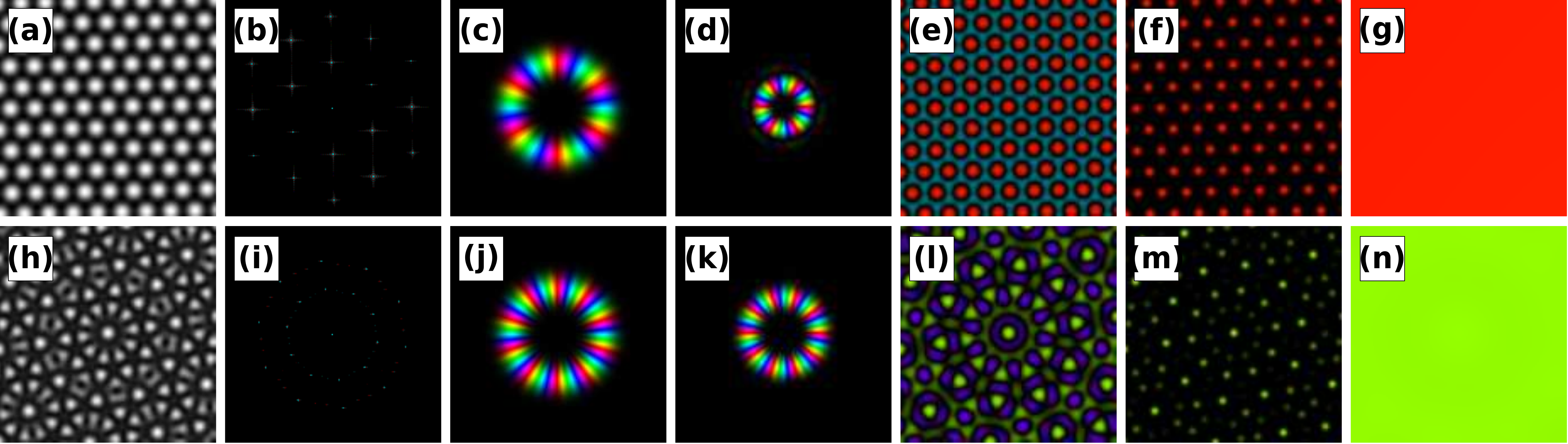}
\caption{The different components of Eq. (\ref{eq-phi}). Amplitude is mapped to brightness and phase to hue (real-valued fields are shown in gray scale). The top row is for a hexagonal crystal and the bottom row for a 10-fold quasicrystal. Panels (a) and (h) give $\psi$ and (b) and (i) the respective Fourier transforms. Panels (c) and (j), and (d) and (k) show $K(k_x,k_y)$ in Fourier space and in direct space, respectively. Note that $K$ is commensurate with the first set of peaks in the spectrum of $\psi$. Panels (e) and (l) present the convolution $\left[ \psi - \min{\left(\psi\right)} \right] \ast K$. Panels (f) and (m) give $\left\lbrace \left[ \psi - \min{\left(\psi\right)} \right] \ast K \right\rbrace \left[ \psi - \min{\left(\psi\right)} \right]$ that eliminates the anti-phase in (e) and (l). Finally, panels (g) and (n) reveal $\phi$, which is uniform for a pristine crystal or quasicrystal.}
\label{fig-orientation}
\end{figure*}

We start with a 2D density field $\psi \equiv \psi(x,y)$, describing a crystalline system, which can be transformed into a smooth, complex-valued orientation field $\phi$ whose argument $\arg{\left(\phi\right)}$ represents the local orientation and whose norm $\left| \phi \right|$ indicates the local crystalline order, or lack thereof, namely defects. The orientation field $\phi$ is given by
\begin{equation}
	\label{eq-phi}
	\phi = \left\lbrace \left[ \left( \psi - \min{\left(\psi\right)} \right) \ast K \right] \left[ \psi - \min{\left(\psi\right)} \right] \right\rbrace \ast G,
\end{equation}
where $\ast$ indicates a convolution and $G$ is a Gaussian kernel just wide enough to filter out the atomic-level structure. The kernel $K$ is given in Fourier space by
\begin{equation}
K \left( \boldsymbol{\mathrm k} \right) = \exp \left[- \left(\left| \boldsymbol{\mathrm k} \right| - q \right)^2/ \left( 2\sigma^2 \right) + i \: m \arg{\left(\boldsymbol{\mathrm k} \right)} \right],
\end{equation}
%
%
%
%
%
%
%
%
where ${\bf \rm k}=(k_x,k_y)$ and $q$ is a wave number corresponding to a characteristic length scale, say the lattice constant ($q = 1$ in our PFC model), $\sigma$ controls the spread of the kernel ($\sigma = 1/5$ appears to work in all the cases here), $i$ is the imaginary unit and $m$ indicates the rotational symmetry of the (quasi-)lattice (respectively, $m = 2, 4, 6$ for stripe, square and hexagonal, or honeycomb, lattices). It appears possible to form $\phi$ for various even-fold symmetric (quasi-)lattices. Odd-fold quasi-lattices display double-fold symmetry centers whose degeneracy leads to $\phi = 0$ in the bulk. Figure \ref{fig-orientation} visualizes the different components of Eq. (\ref{eq-phi}) for a hexagonal crystal and a 10-fold quasicrystal.

\begin{figure*}
\includegraphics[width=\textwidth]{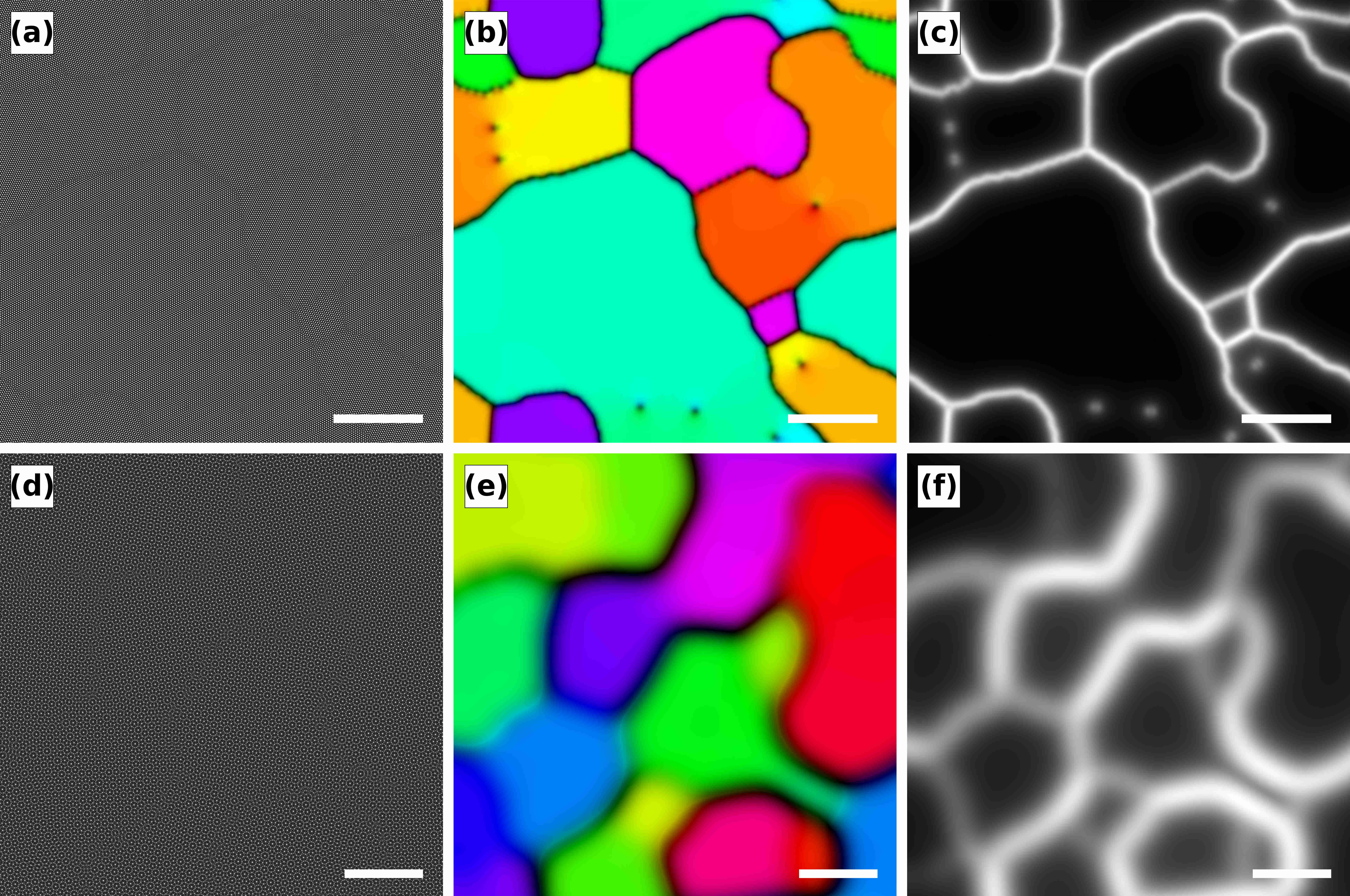}
\caption{Polycrystalline hexagonal (top panels) and 12-fold quasicrystal (bottom panels) from the PFC model. In (a) and (d) we show the density field $\psi$, in (b) and (e), the orientation field $\phi$, and in (c) and (f), the deformation field $\chi$. The scale bars have lengths $40 a_\textrm{hex}$ and $20 a_\textrm{hex}$ for the hexagonal and the 12-fold systems, respectively.}
\label{fig-poly}
\end{figure*}

The defects and changes in the crystallographic orientation can be mapped by the magnitude of the gradient of the orientation field
\begin{equation}
	\left| \nabla \phi \right| = \sqrt{ \Re \left( \phi_x \right)^2 + \Im \left( \phi_x \right)^2 + \Re \left( \phi_y \right)^2 + \Im \left( \phi_y \right)^2 },
\end{equation}
where $\Re$ and $\Im$ give the real and imaginary parts, respectively, while $\phi_x$ and $\phi_y$ denote the partial derivatives of $\phi$ with respect to the $x$ and $y$ directions. From the gradient, one can construct a filtered, smooth deformation field $\chi$ as
\begin{equation}
\label{eq-chi}
	\chi = \sum_{n = 0}^{2^n a < \min \left( W, H \right)/2} \frac{ \left| \nabla \phi \right|^p \ast \exp \left[ - \left| \boldsymbol{\mathrm r} \right|^2 / \left( 2 \cdot 2^{2n} a^2 \right) \right] }{ \max \left\lbrace \left| \nabla \phi \right|^p \ast \exp \left[ - \left| \boldsymbol{\mathrm r} \right|^2 / \left( 2 \cdot 2^{2n} a^2 \right) \right] \right\rbrace },
\end{equation}
where $a$ is the lattice constant, $W$ and $H$ are the dimensions of the system, and $p$ is a tunable exponent (to be discussed in detail in Sec. \ref{sec-assessment}). The right-hand side of Eq. (\ref{eq-chi}) gives a sum of normalized convolutions between a power of the gradient and Gaussian kernels of width $2^n$ lattice constants. The sum is truncated before the kernel width reaches the smaller of the system dimensions. Figure \ref{fig-poly} demonstrates $\phi$ and $\chi$ for polycrystalline systems of hexagonal and 12-fold quasicrystalline lattice types.

%
%
%
%
%
%
%

\begin{figure*}
\includegraphics[width=\textwidth]{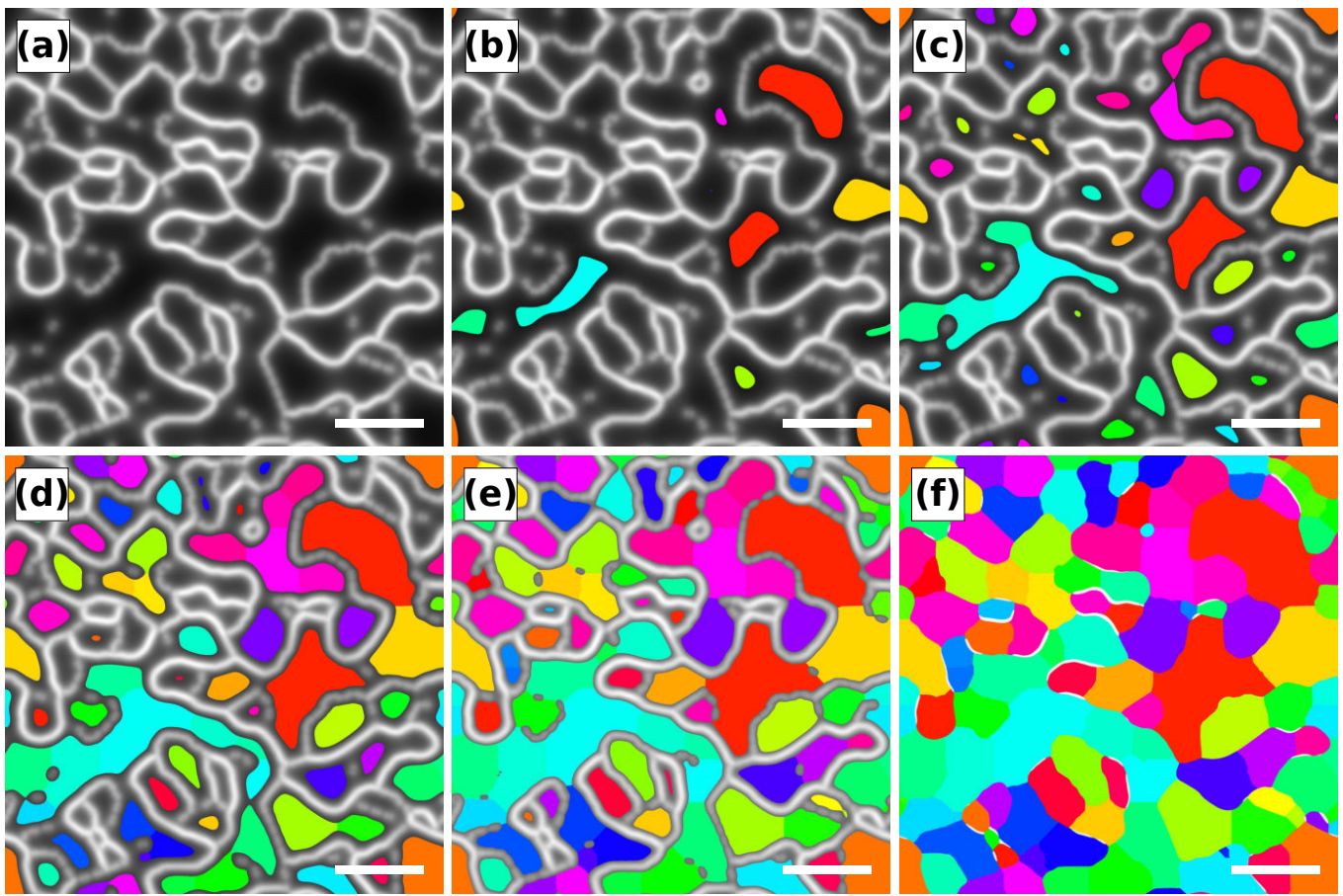}
\caption{Time series of subdomain growth in $\chi$. The panels (a)-(f) show snapshots of the growth procedure. The scale bars have length $40 a_\textrm{hex}$.
}
\label{fig-growth}
\end{figure*}

With the help of $\chi$, a polycrystalline system can be segmented into subdomains. Each local minimum in $\chi$ is treated as a seed and the subdomains are grown from these seeds by climbing the $\chi$ value landscape. All growth fronts climb $\chi$ at the same rate and stop when the subdomains collide. The lattice orientation of a subdomain is given by the average of $\phi$ over it. This procedure is illustrated by the time series in Fig. \ref{fig-growth}. We considered using $\left| \Delta \phi \right|^p$ directly as the deformation field in lieu of $\chi$, but the former has a large number of local minima that greatly exceeds the number of real grains. This brings about much additional complexity and, ultimately, leads to a failure to properly detect the grains. Filtering $\left| \Delta \phi \right|^p$ further using a single Gaussian kernel is also not ideal, as there is a trade-off between getting rid of excess local minima and smoothing out small-scale features of the microstructure. On the other hand, while $\chi$ is very smooth far from defects and displays fewer excess local minima, it still captures the paths of the grain boundaries and the positions of isolated dislocations accurately.

\begin{figure*}
\includegraphics[width=\textwidth]{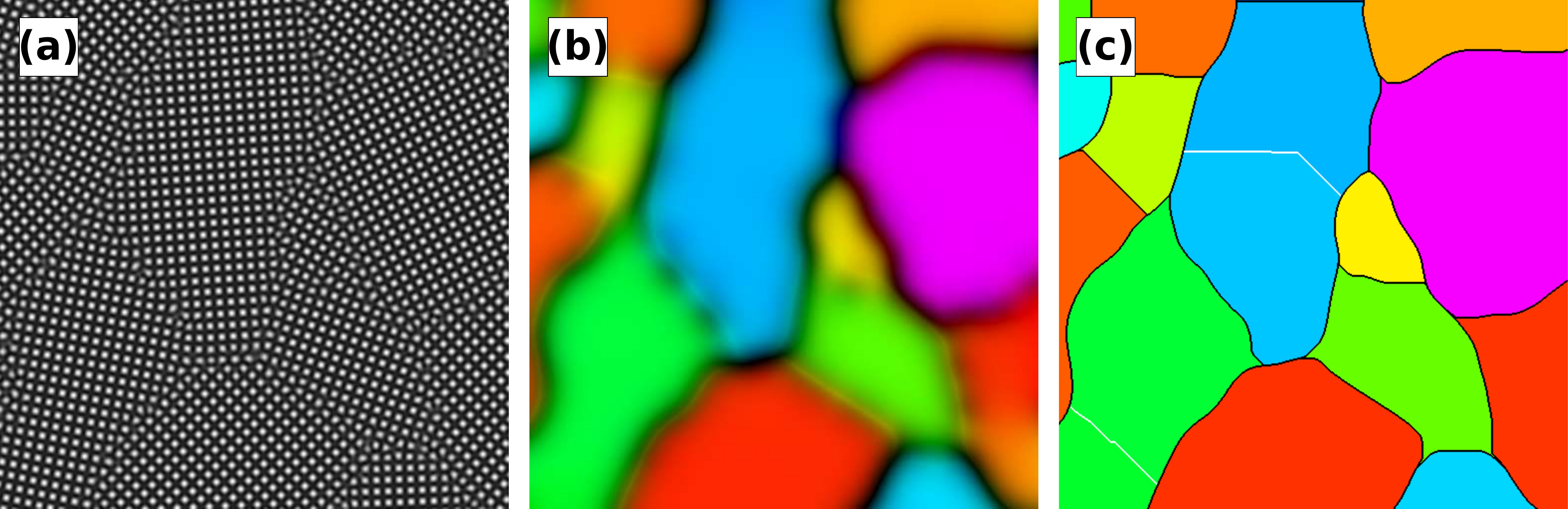}
\caption{Example of the subdomain merging step for a square system. (a) The density field $\psi$, (b) the orientation field $\phi$ and (c) the grain structure extracted. White borders in (c) indicate that the neighboring subdomains have been merged as parts of the same grain, whereas black borders correspond to grain boundaries between the grains that have been identified
as different.}
\label{fig-merging}
\end{figure*}

Quite often some grains contain multiple local minima of $\chi$ and are consequently subdivided into multiple subdomains. As a final step of the grain extraction algorithm, neighboring subdomains are merged if they satisfy certain conditions. Various criteria were considered but a simple misorientation-based criterion was found to be sufficient: merge neighboring subdomains if the relative difference between the two lattice orientations $\theta < \theta^\ast$. The optimal choice of $\theta^\ast$ for each lattice type is discussed in Sec. \ref{sec-assessment}. Figure \ref{fig-merging} gives an example of the subdomain merging step where the growth step has resulted in two grains (green and blue), both subdivided into two subdomains (b) and the subdomains have then been merged together (c) to recover the true grains. An additional condition was introduced for very small grains below a certain linear size: such grains are merged with the neighbor that is closest in lattice orientation. As the limit a linear size of 5 times the lattice constant was used. Such grains are just barely larger than the dislocations enclosing them. All lattice types considered in this work display roughly similar length scales whereby the approximate dimensionless lattice constant for the hexagonal lattice $a_\textrm{hex} = 4\pi/\sqrt{3} \approx 7.3$ was used for all of them. 

Regarding the computational cost of the method, its two bottlenecks are computing the deformation field $\chi$ and the subdomain growth step. We implemented the several convolutions in the former using parallelized fast Fourier transforms. The latter was realized as a serial iterative algorithm due to its complexity. We expect that the latter step can be sped up significantly by using a better, parallelized algorithm. It takes on the order of a few minutes for a quad-core desktop PC to fully process a system of 8192 $\times$ 8192 grid points. The computational performance of the method is discussed in more detail in Sec. \ref{sec-performance} in the Appendix.

\subsection{Model systems}
\label{sec-methods-model-systems}

We applied the grain extraction method to study the microstructure and its evolution in polycrystalline systems of different lattice types. We considered regular square and hexagonal lattices, as well as 10- and 12-fold quasicrystalline ones. Random polycrystalline 2D systems were obtained from large-scale grain coarsening simulations carried out using a two-mode PFC model. PFC models are a family of continuum methods for structural and elastoplastic modeling of crystalline matter at the atomistic scale. The main advantage with PFC models is the access to long, diffusive time scales over which microstructure evolution takes place. Mesoscopic systems can be handled readily with atomic-level resolution. Systems modeled using PFC are described in terms of smooth, classical density fields $\psi$ whose evolution is governed by a free energy functional \cite{ref-pfc-debut, ref-ken-martin, ref-nik-ken}. We used a two-mode PFC free energy functional \cite{ref-pfc-fcc,ref-cristian-growth-modes}
\begin{equation}
\label{eq-functional}
F = \int d\boldsymbol{r} \left( \frac{\psi}{2} \left( R + \prod_{n=1}^{N} \left( q_n^2 + \nabla^2 \right)^2 \right) \psi + \frac{\psi^4}{4}\right),
\end{equation}
where $R$ is related to temperature,
$N =$ 1 or 2 indicates the number of modes, and the wave numbers $q_n$ control the periodic length scales in $\psi$. We evolved $\psi$ forward in time assuming diffusive dynamics as
\begin{equation}
\label{eq-dynamics}
\frac{\partial \psi}{\partial t} = \nabla^2 \frac{\delta F}{\delta \psi},
\end{equation}
where $\delta/\delta \psi$ indicates a functional derivative with respect to $\psi$. Diffusive dynamics strictly conserve the average density $\bar{\psi}$ that, together with $R$ and $q_n$, controls the lattice type. The average densities and model parameters used for the four different lattice types considered in this work are given in Table \ref{tab-parameters}. The average densities and model parameters for the hexagonal lattice and the quasicrystals were adopted from Refs. \cite{ref-backofen} and \cite{ref-cristian-growth-modes}, respectively, whereas for the square lattice they were found by trial and error. We used the semi-implicit spectral method given in Ref. \cite{ref-nik-ken} although similar spectral methods have been used elsewhere in the literature, for example Ref. \cite{ref-elsey-numerics} used by Refs. \cite{ref-backofen, ref-gabriel}. The specific numerical method and parameters do not appreciably influence the grain extraction algorithm since the precise atomic behavior is washed out in computing the orientation field. Note that periodic boundary conditions ensue from the use of a spectral method.

\begin{table}
\centering
\caption{Model parameters and average densities used for the four different lattice types considered in this work. The first column, titled $m$, indicates the lattice types by their rotational symmetry.}
\label{tab-parameters}
\begin{tabular}{lccccc}
\hline
\hline
$m$ & $R$ & $N$ & $q_1$ & $q_2$ & $\bar{\psi}$ \\
\hline
4 & $-0.18$ & 2 & 1 & $\sqrt{2}$ & $-0.28$ \\
6 & $-0.18$ & 1 & 1 & -- & $-0.25$ \\
10 & $-0.07$ & 2 & 1 & $2\cos{\left( \pi/5 \right)}$ & $-0.161060$ \\
12 & $-0.25$ & 2 & 1 & $2\cos{\left( \pi/12 \right)}$ & $-0.314904$ \\
\hline
\hline
\end{tabular}
\end{table}

\begin{figure}
\centering
\includegraphics[width=0.333\textwidth]{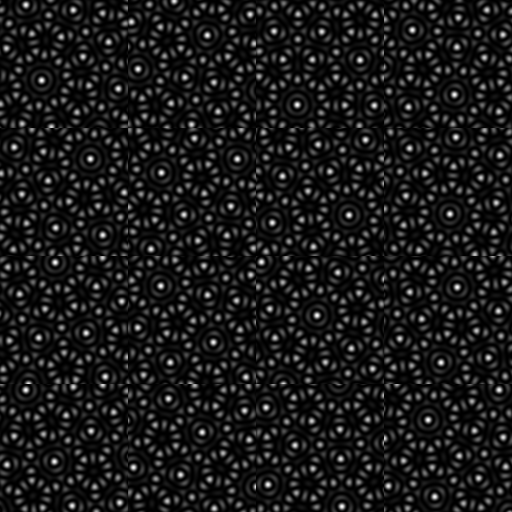}
\caption{Tiled, random initial state for a grain coarsening simulation. This blow-up demonstrates 4 $\times$ 4 square tiles of 10-fold quasicrystal in random orientations. The quasicrystalline structure is obtained from a sum of plane waves.}
\label{fig-tiles}
\end{figure}

For the original PFC model with $N=1$, for parameters where the hexagonal phase has the lowest energy, the liquid phase is always linearly unstable with respect to small deformations \cite{ref-cheng-algorithm}. Consequently, in order to grow a polycrystalline configuration from a liquid initial state,  for most parameter choices it is sufficient to start with a random density field. However, for the quasicrystal systems modeled using Eq. (\ref{eq-functional}) with $N = 2$ and the parameters given in Table \ref{tab-parameters}, the liquid is linearly stable to small perturbations. The critical size of initial seeds for stable growth is relatively large for the present quasi-lattices with the model parameters and the average densities chosen \cite{ref-cristian-growth-modes,ref-cristian-qc-interfaces}. Stability of the quasicrystalline phases was ensured by exploiting initial states with moderate-sized square tiles of the lattice type desired in random orientations as in Fig. \ref{fig-tiles}. All initial lattice structures were obtained with one-mode approximations, \textit{i.e.} by summing plane waves \cite{ref-plane-waves}.

The method was also tested on molecular dynamics (MD) generated data of free-standing polycrystalline monolayer graphene to investigate the impact of thermal fluctuations -- giving rise to displacements of atoms and to out-of-plane buckling of the sheet -- on the performance of the method. First, relaxed PFC density fields for polycrystalline graphene were converted into sets of atomic coordinates. The approx. 48 $\times$ 48 nm$^2$ systems were thermalized at both 1 K and 300 K using a GPUMD code \cite{ref-gpumd,ref-gpumd-github} with the Tersoff potential \cite{ref-tersoff, ref-lindsay-broido}. Here, we reused systems from our previous work on thermal transport in polycrystalline graphene \cite{ref-bimodal} and the details of the PFC and MD simulations can be found there in full. The relaxed MD coordinates were converted back into 2D density fields suitable for the present grain extraction code by first projecting them onto the $xy$ plane and smoothing atoms with Gaussian peaks.

\section{Results}

\subsection{Assessment of the grain extraction method}
\label{sec-assessment}

This subsection is dedicated to the assessment of the performance of the grain extraction method and to its optimization to reproduce the hand segmentations of the authors of the patched network of grains in a polycrystalline system. The preliminary networks of subdomains are first investigated, before optimizing subdomain merging step to match human judgment. Lastly, the method's applicability to molecular dynamics data is demonstrated.

\subsubsection{Assessment of the subdomain network}

\begin{figure*}
\includegraphics[width=\textwidth]{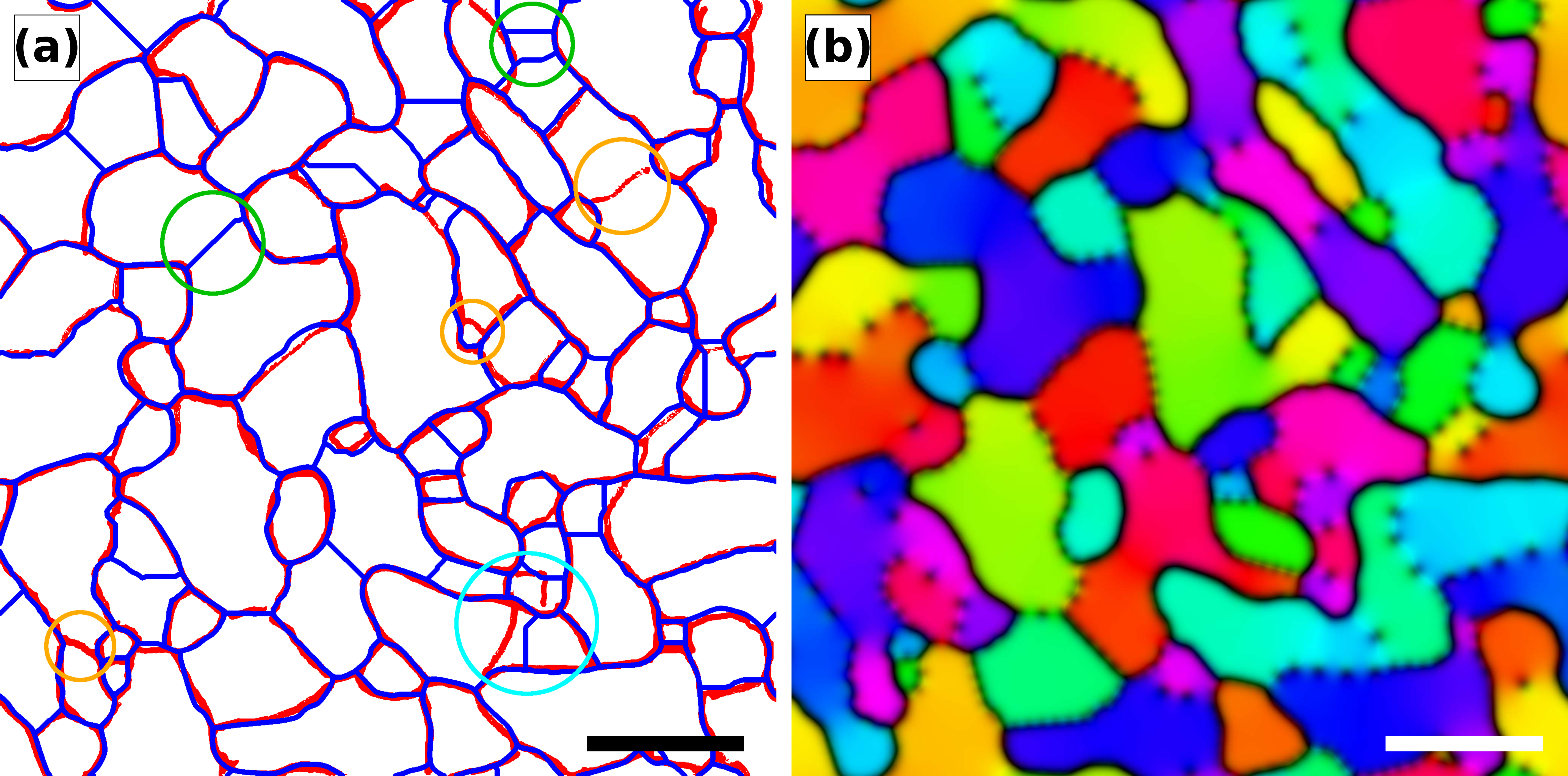}
\caption{(a) A comparison between the grains determined by a visual inspection (red) and the subdomains produced by the present method (blue) for a hexagonal system. The green, cyan and orange circles indicate subdomain borders, minor differences in grain boundary delineation, and inconsistencies, respectively. (b) The corresponding orientation field $\phi$. The scale bars have length $40 a_\textrm{hex}$. See text for details.}
\label{fig-K-grains-vs-subdomains}
\end{figure*}

A prerequisite for capturing the correct grain structure is a patchwork of subdomains that captures the outlines of the grains. Figure \ref{fig-K-grains-vs-subdomains} demonstrates in red color the grain boundaries in a polycrystalline system as determined by one of the authors here (KRE) by a simple visual examination of the atomic number density map. The blue lines are the corresponding subdomain boundaries determined by the present method. The most typical difference between the two are the subdomain borders inside the grains due to excess local minima; a few examples have been highlighted in green. These are not a major issue as long as the subdomains are merged appropriately. Minor differences in grain boundary delineation, highlighted in cyan, are another fairly typical and rather unimportant feature. Our numerical method misses some boundaries proposed by KRE, highlighted in orange, but these most often correspond to grain boundaries whose existence is somewhat amnbiguous.

\begin{figure*}
\includegraphics[width=\textwidth]{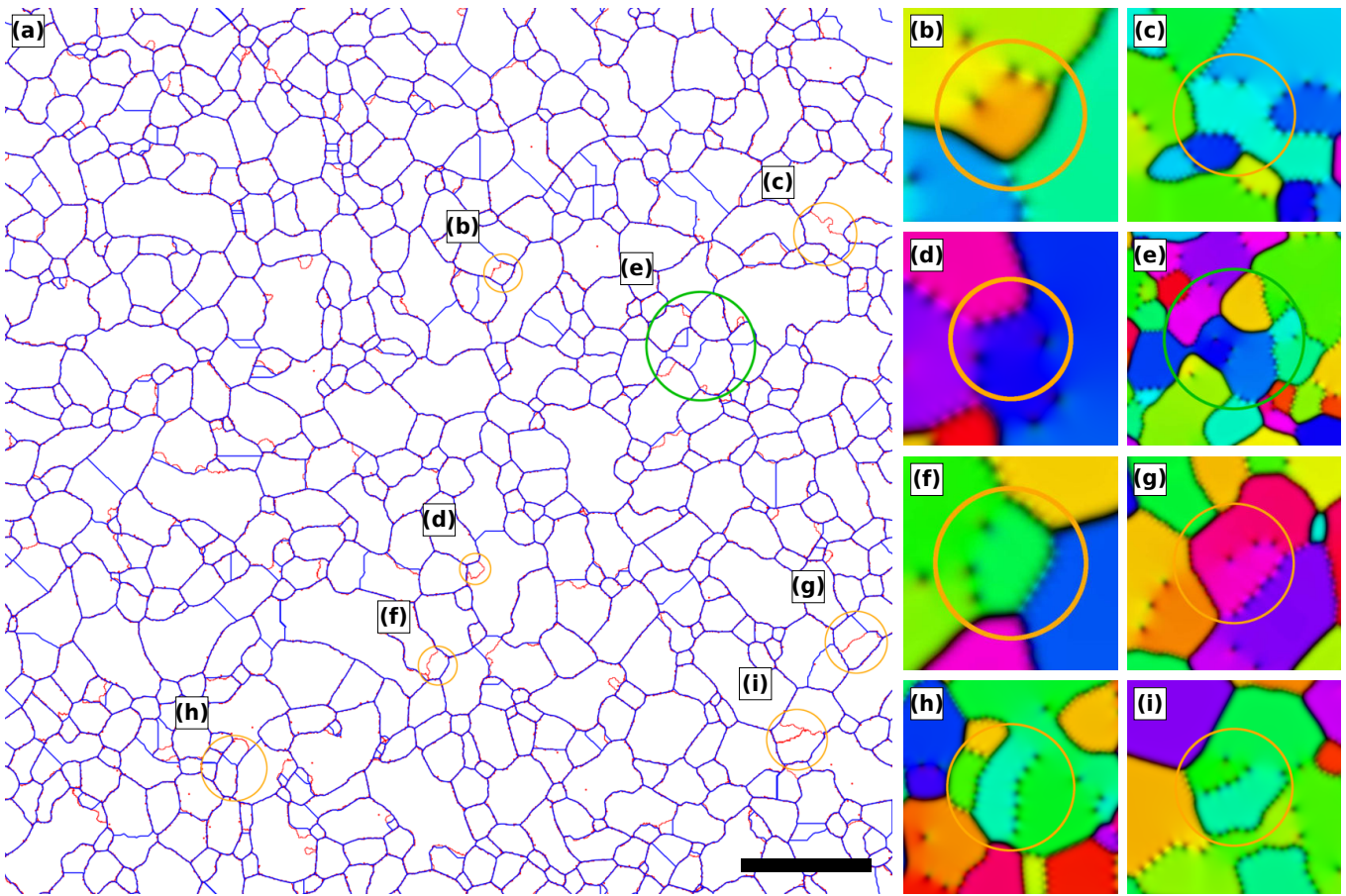}
\caption{(a) A comparison between the grains and subdomains produced by the previous \cite{ref-gabriel} (red) and the present (blue) methods, respectively. (b)-(i) Blow-ups of the orientation field $\phi$ around certain structures (indicated by their respective labels in (a)). The scale bar has length $150 a_\textrm{hex}$. The blow-ups are not to scale; compare with (a).
See text for details.}
\label{fig-Gc-vs-Pc}
\end{figure*}

We compared the method to a previous atom-based method from Ref. \cite{ref-gabriel}. The previous method is applicable to hexagonal lattices and has been shown to be robust and highly accurate. Figure \ref{fig-Gc-vs-Pc} offers a comparison between the grains and the subdomains given respectively by the previous (red) and the present method (blue). The overall agreement between the two methods is very good and most deviations involve minor differences in grain delineation and small potential artifacts due to ambiguous grain boundaries and individual dislocations creeping close to grain boundaries; some examples are highlighted within the green circle. There is a handful of more complicated structures, circled in orange, where the present method may misplace or miss ambiguous grain boundaries. As discussed in Ref. \cite{ref-gabriel}, such boundaries are very difficult to recover in a robust fashion, either with manual or numerical segmentations. Ultimately, such problems concern only about 1\% of all the grains in the system.

\subsubsection{Assessment and optimization of subdomain merging}

The final grain structures obtained from the subdomain merging step were benchmarked and optimized against hand segmentations of grain network images. The hand segmentations were generated by first plotting the subdomains given by the method. Authors PH, KRE and GMLaB then used image manipulation software to recolor the subdomains, using (non-)identical colors for two neighboring subdomains to indicate that they should (not) be merged. The manipulated images were loaded into the grain extraction program and the code merged the subdomains accordingly for further analysis.

For each lattice type, multiple systems at different time steps and with different average grain sizes were considered. A more comprehensive assessment was carried out for hexagonal systems for which PH, KRE and GMLaB all prepared their own hand segmentations. The agreement between the segmentations of the code and those of the authors was measured by calculating the fraction of neighboring subdomain pairs that were treated, \textit{i.e.}, merged or not merged, similarly. The misorientation limit $\theta^\ast$ for the merging criterion was varied to find the optimal value for each lattice type. For the norm of the gradient $|\nabla \phi|^p$ in Eq. (\ref{eq-chi}), $p = 2$ for the periodic lattices and $p = 1$ for the quasi-lattices appeared to increase the extraction accuracy.

\begin{figure}
\centering
\includegraphics[width=0.5\textwidth]{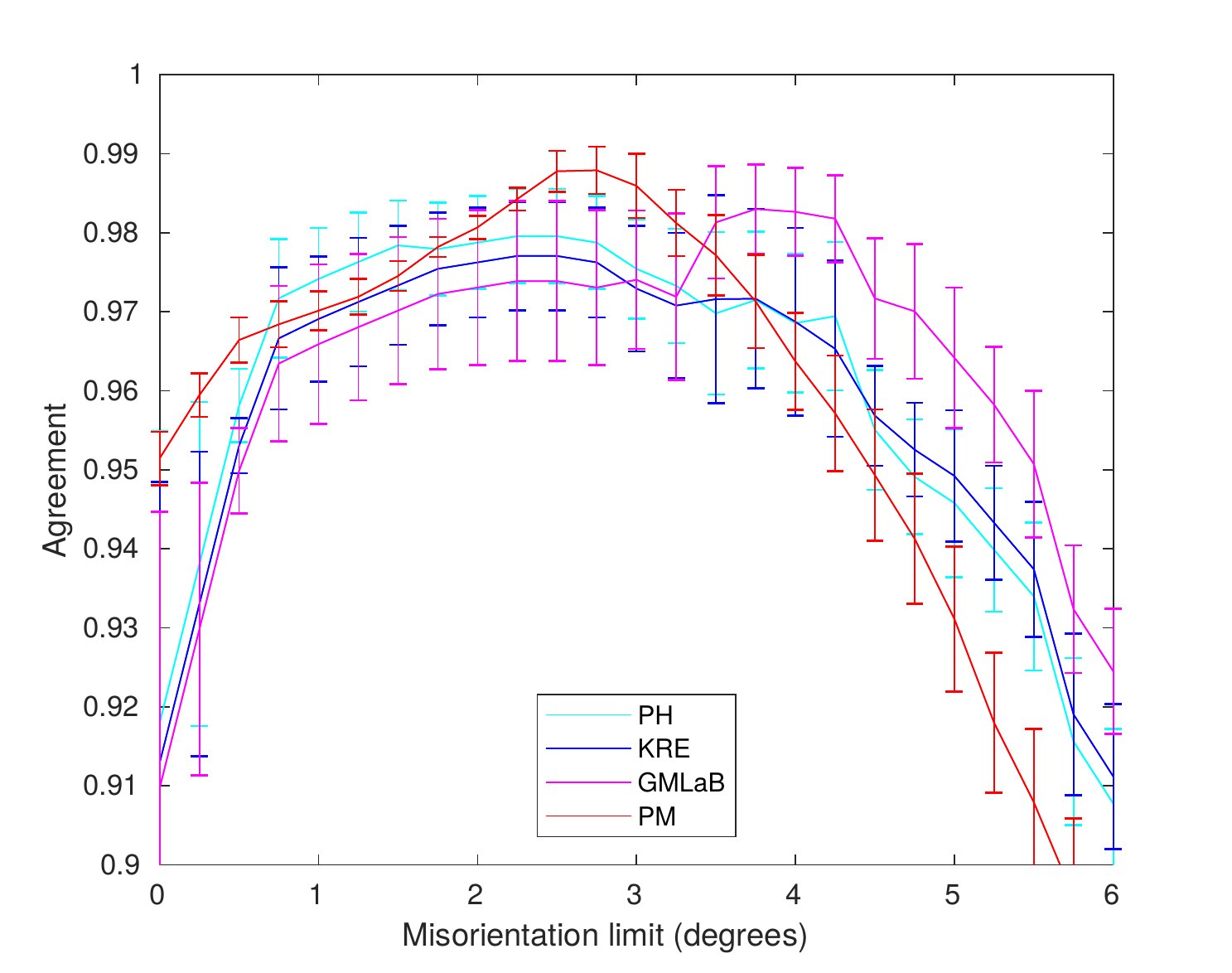}
\caption{
Level of agreement (normalized to a maximum level of unity) of the present method with the hand segmentations of the authors PH, KRE and GMLaB, and with the segmentations of the previous method (PM) \cite{ref-gabriel}, for hexagonal systems, as a function of the misorientation limit $\theta^\ast$.}
\label{fig-everybody-vs-everybody}
\end{figure}

Figure \ref{fig-everybody-vs-everybody} 
compares the level of agreement of the present method with the manual segmentations and with the previous atom-based method, for hexagonal systems as a function of $\theta^\ast$. Note that the limit $\theta^\ast = 0^\circ$ corresponds to omitting the subdomain merging step and treating each subdomain as a separate grain. The five hexagonal systems hand segmented had 622 pairs of neighboring subdomains in total and their average linear grain sizes varied from $\approx 180$ to $\approx 590$ (in dimensionless units where the approximate lattice constant is $a_\textrm{hex} = 4 \pi / \sqrt{3} \approx 7.3$). The average linear grain size is given by
\begin{equation}
\left\langle d \right\rangle = \sqrt{S/N},
\end{equation}
where $S$ is the total area of a system and $N$ is the number of grains in it. Comparison to the previous method was carried out similarly to the hand segmentations by comparing the colors in the image files representing the numerical segmentation. The segmentations of the previous method were prepared using the fixed set of parameters found optimal in Ref. \cite{ref-gabriel}. The five much larger hexagonal systems segmented in an automated fashion by the previous method had a total of 13673 pairs of neighboring subdomains and the average linear grain sizes varied from $\approx 170$ to $\approx 660$. The values and the error bars shown are the average and standard error, respectively, of the agreements for the individual systems segmented.

Figure \ref{fig-everybody-vs-everybody} 
shows that although the error margins are relatively large at the scale shown, the present method performs very well as compared to the hand segmentations of both PH and KRE, peaking around $\theta^\ast \approx 2.5^\circ$. The agreement with GMLaB's hand segmentations appears slightly higher for $\theta^\ast > 3^\circ$ and peaks around $\theta^\ast \approx 3.75^\circ$. The agreement with the previous method's segmentation is a bit lower for $\theta^\ast > 4^\circ$, a bit higher for $\theta^\ast < 0.5^\circ$ and peaks around $\theta^\ast \approx 2.75^\circ$. Despite these minor differences, the present method's agreement with all segmentations is high and consistent for the wide, approximate range of $1^\circ \leq \theta^\ast \leq 4^\circ$. While the grain boundaries with such low misorientation are often somewhat ambiguous, all manual and the two numerical segmentations are mutually consistent. This shows that the two grain extraction methods could be substituted for the extremely tedious manual segmentation with little or no loss in accuracy. Table \ref{tab-segmentation-hex} summarizes the maximal agreement and the corresponding $\theta^\ast$ for each author.

\begin{table}
\centering
\caption{Maximal level of agreement and the corresponding $\theta^\ast$ of the present method with the hand segmentations of PH, KRE and GMLaB, and with the segmentations of the previous method \cite{ref-gabriel}, all for hexagonal systems.}
\label{tab-segmentation-hex}
\begin{tabular}{lcc}
\hline
\hline
Segmentation & Agreement & $\theta^\ast$ (degrees) \\
\hline
PH & 0.980 $\pm$ 0.006 & 2.5 \\
KRE & 0.977 $\pm$ 0.007 & 2.5 \\
GMLaB & 0.983 $\pm$ 0.006 & 3.75 \\
previous method & 0.988 $\pm$ 0.003 & 2.75 \\
\hline
\hline
\end{tabular}
\end{table}

\begin{figure*}
\includegraphics[width=\textwidth]{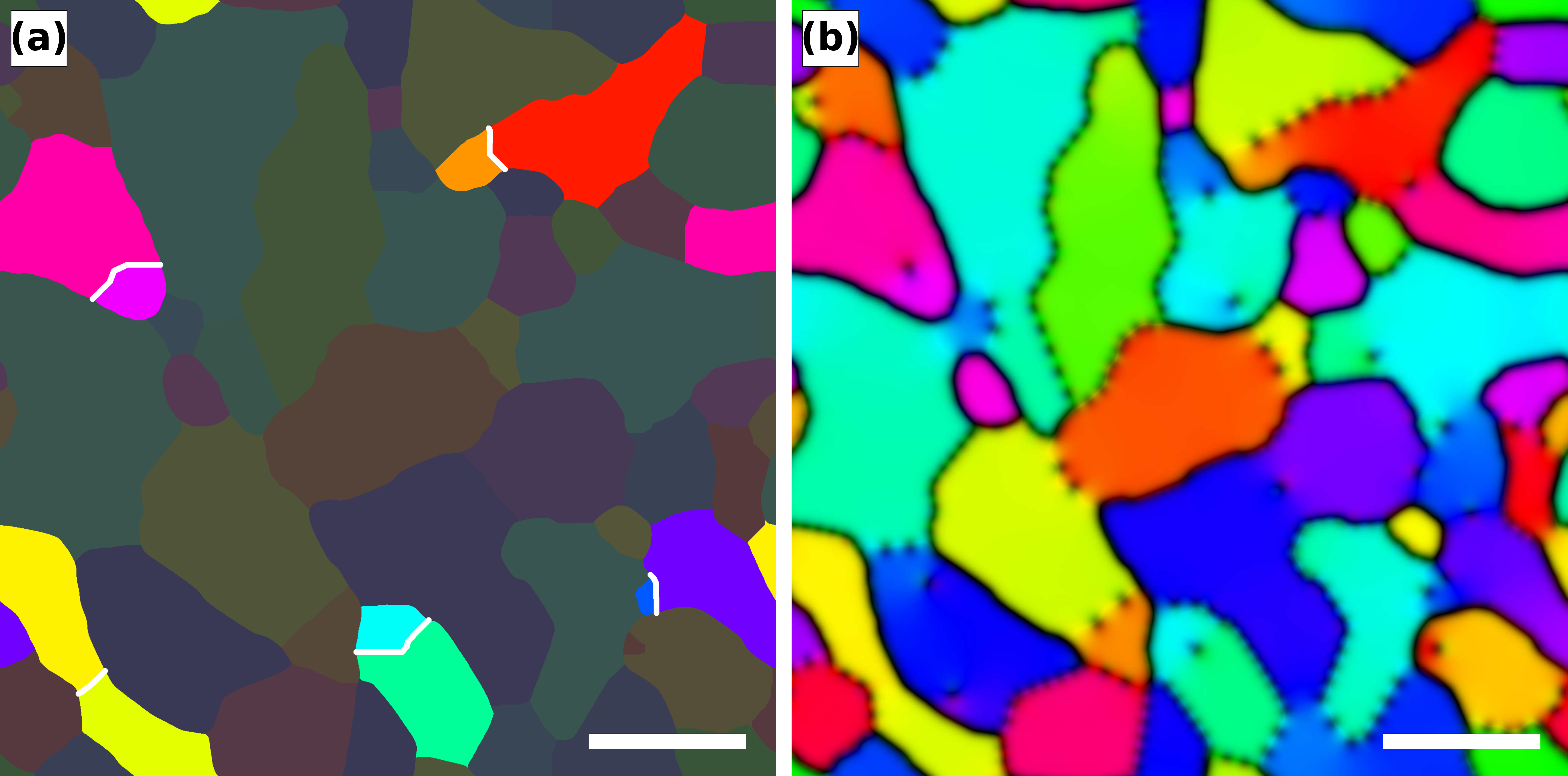}
\caption{(a) Comparison between the hand segmentations of authors PH and GMLaB. Pairs of subdomains that were merged differently in the two hand segmentations, as well as their mutual interfaces have been highlighted. (b) The corresponding orientation field $\phi$. The scale bars have length $40 a_\textrm{hex}$.}
\label{fig-G-vs-P}
\end{figure*}

Based on Fig. \ref{fig-everybody-vs-everybody}, the hand segmentations appear somewhat different between the authors PH and GMLaB, as well as between KRE and GMLaB. Figure \ref{fig-G-vs-P} shows typical examples of disagreement between the hand segmentations of PH and GMLaB. The pairs of subdomains treated differently, and their mutual interfaces, have been highlighted. Most cases involve corners or appendages of grains where there is some change in orientation and individual dislocations are also often involved. All cases of disagreement are typically somewhat ambiguous.

\begin{figure}
\centering
\includegraphics[width=0.5\textwidth]{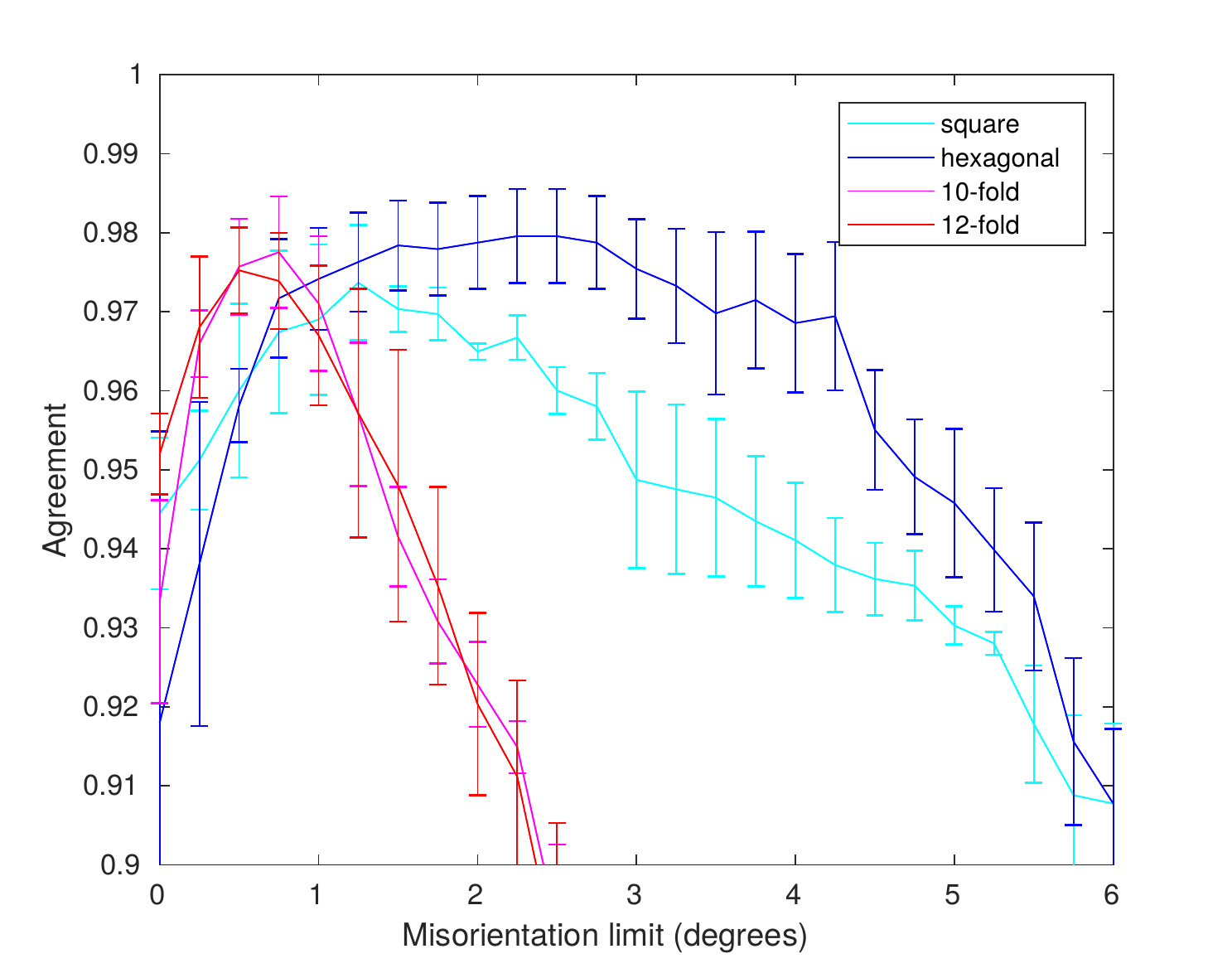}
\caption{Level of agreement of the present method with the hand segmentations of PH for square and hexagonal, as well as for 10- and 12-fold quasicrystalline lattices, as a function of $\theta^\ast$.}
\label{fig-4-6-10-12-fold}
\end{figure}

Figure \ref{fig-4-6-10-12-fold} demonstrates the present method's agreement with the hand segmentations of PH for all four lattice types considered in this work as a function of $\theta^\ast$. For the hexagonal lattice, the same data set as in Fig. \ref{fig-everybody-vs-everybody} is shown, but to reiterate, the maximal level of agreement for the hexagonal lattice is 0.980 $\pm$ 0.006 at $\theta^\ast \approx 2.5^\circ$. For the square lattice, the agreement is maximized at $\theta^\ast \approx 1.25^\circ$ and is 0.974 $\pm$ 0.007.  For the 10- and 12-fold quasicrystals, the agreement is maximized at $\theta^\ast \approx 0.75^\circ$ and 0.5$^\circ$, and is 0.978 $\pm$ 0.007 and 0.975 $\pm$ 0.005, respectively. Compared to the periodic lattices, the respective agreements are much more sensitive to $\theta^\ast$, as the agreement falls below 0.9 already where $\theta^\ast \geq 2.5^\circ$. We would like to point out that the optimal value for $\theta^\ast$ need not be proportional to the order of the rotational symmetry $m$. The present method shows a varying tendency to produce excess subdomains for the different lattice types and, the more subdomains there are, the smaller the misorientation between them, and \textit{vice versa}. The tendency to subdivide grains into subdomains depends on the spread of the Gaussian smoothing kernel $G$ in Eq. (\ref{eq-phi}), required to filter out the atomic-level structure, and on the exponent $p$ in Eq. (\ref{eq-chi}). Table \ref{tab-segmentation-others} summarizes the maximal levels of agreement and the corresponding $\theta^\ast$ for each of the three other lattice types.
\begin{table}
\centering
\caption{Maximal level of agreement and corresponding $\theta^\ast$ of the present method with the hand segmentations of PH for square, and 10- and 12-fold quasicrystalline systems. The number of neighboring subdomain pairs $P$ and the ranges of the average linear grain sizes $\langle d \rangle$ are also given.}
\label{tab-segmentation-others}
\begin{tabular}{lcccc}
\hline
\hline
Lattice type & Agreement & $\theta^\ast$ (degrees) & P & $\langle d \rangle$ \\
\hline
square & 0.974 $\pm$ 0.007 & 1.25 & 1496 & $280-650$ \\
10-fold & 0.978 $\pm$ 0.007 & 0.75 & 1297 & $290-530$\\
12-fold & 0.975 $\pm$ 0.005 & 0.5 & 1031 & $270-580$ \\
\hline
\hline
\end{tabular}
\end{table}
\subsubsection{Applicability to molecular dynamics data}
\begin{figure*}
\includegraphics[width=\textwidth]{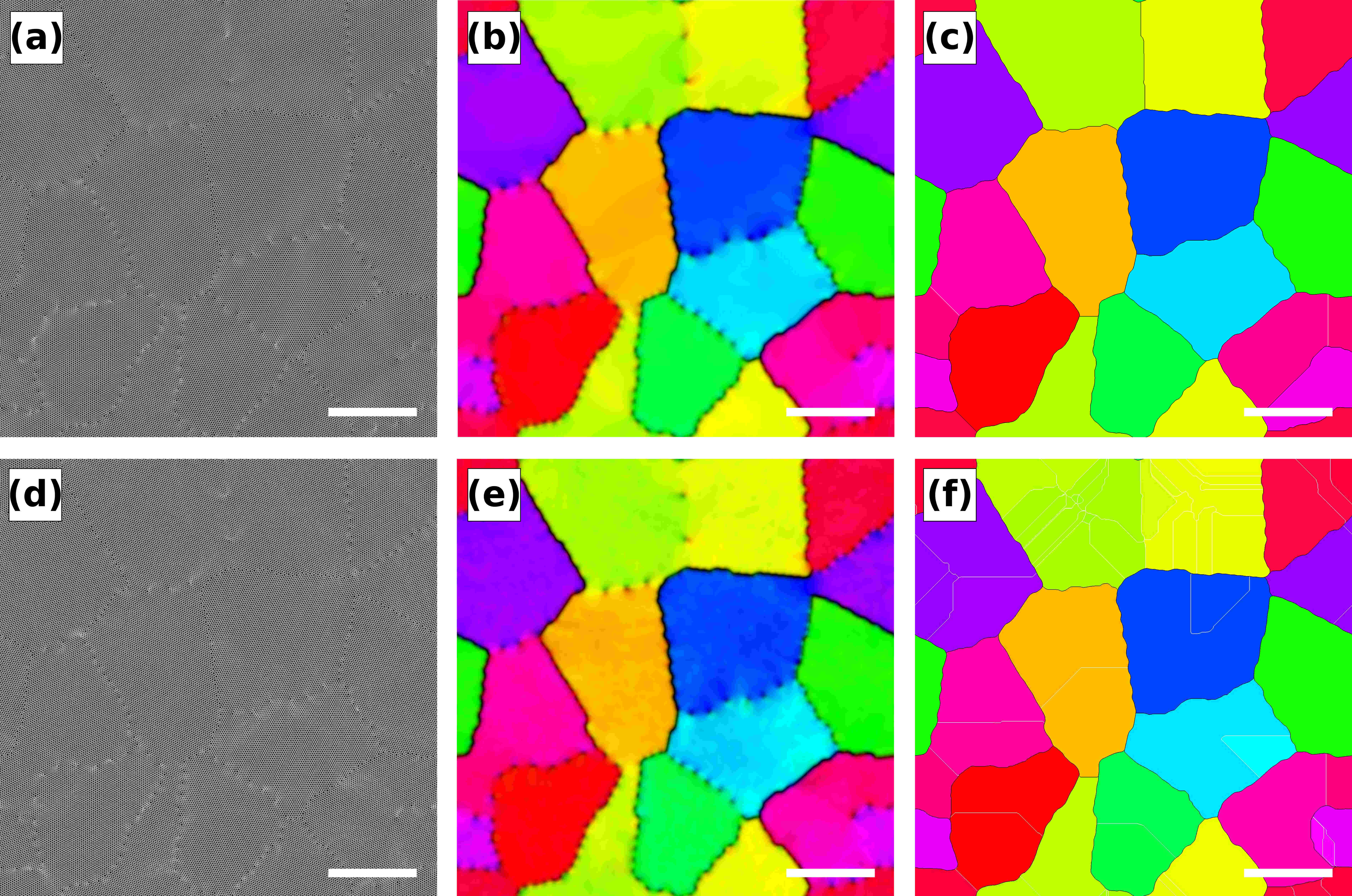}
\caption{Demonstration of the grain extraction method for MD data on graphene. The top (bottom) row corresponds to a system thermalized at 1 K (300 K). Panels (a) and (d) give the density field obtained by substituting small Gaussian peaks at the projected atom positions, (b) and (e) give the orientation field $\phi$, and (c) and (f) give the grain structure extracted. In (c) and (f), white borders indicate that the neighboring subdomains have been merged as parts of the same grain, whereas black borders give the grain boundaries between true grains. The scale bars have an approximate length of 40 graphene lattice constants.}
\label{fig-MD}
\end{figure*}

Last, Fig. \ref{fig-MD} demonstrates the applicability of the method to MD atomic number density data for graphene. We observed similar results for all samples and showcase here a single example. The 1 K configuration displays faint long-wavelength ripples, due to out-of-plane buckling of the monolayer, but this causes no issues. The thermal fluctuations far greater in the 300 K configuration lead to noticeable short-wavelength ripples in the corresponding orientation field, which results in a multitude of excessive subdomains. Despite this, the method is ultimately able to recover most of the grain structure. Here, $\theta^\ast = 2^\circ$ was used.

At 300 K, the method ends up merging -- erroneously to our opinion -- the two grains at the top of the figure. At 1 K, the dumbbell-shaped composite of two subdomains at the periodic corner of the figure is treated as two separate grains as their misorientation $\theta > \theta^\ast = 2^\circ$. At 300 K, the method considers the corresponding set of subdomains a single grain. Another noticeable difference between the high and the low temperature configurations is the delineation between the grains in lower right, but this case is a somewhat ambiguous one.

\subsection{Microstructural analysis of different lattice types}
\label{sec-analysis}

The present grain extraction method was used to analyze the microstructure and evolution of four different lattice types. Regular periodic square and hexagonal lattices as well as 10- and 12-fold quasi-lattices were studied to compare and to shed light on the microstructure especially in quasicrystals. Various microstructural properties were investigated, but to focus on the most relevant results and to keep this section concise, part of the results are given in detail in Sec. \ref{sec-microstructure-further} in the Appendix. A brief summary of these results is given here.

All values and error bars plotted here are the mean and the standard error, respectively, of parallel realizations of model systems. Unless stated otherwise, results for the lattice types will be listed in the order of increasing $m$: square, hexagonal, 10-fold and 12-fold. For the four lattice types, 16, 16, 32 and 32 parallel realizations of PFC grain coarsening simulations were conducted. All realizations had a size of $8192 \times 8192$ grid points and the spatial discretizations were $\Delta x = \Delta y =$ 0.55, 0.8, 0.5 and 0.4. The systems were evolved for $5 \times 10^6$ time steps each and the time step sizes were $ \Delta t =0.5, 0.4, 0.02, 0.01$. We also compared our systems to random Voronoi tessellations in some instances. A total of 100 random seed points was sampled into each periodic Voronoi system of $4096 \times 4096$ grid points. A total of 1000 parallel realizations were generated.

\subsubsection{Evolution of average linear grain size}

\begin{figure}
\centering
\includegraphics[width=0.5\textwidth]{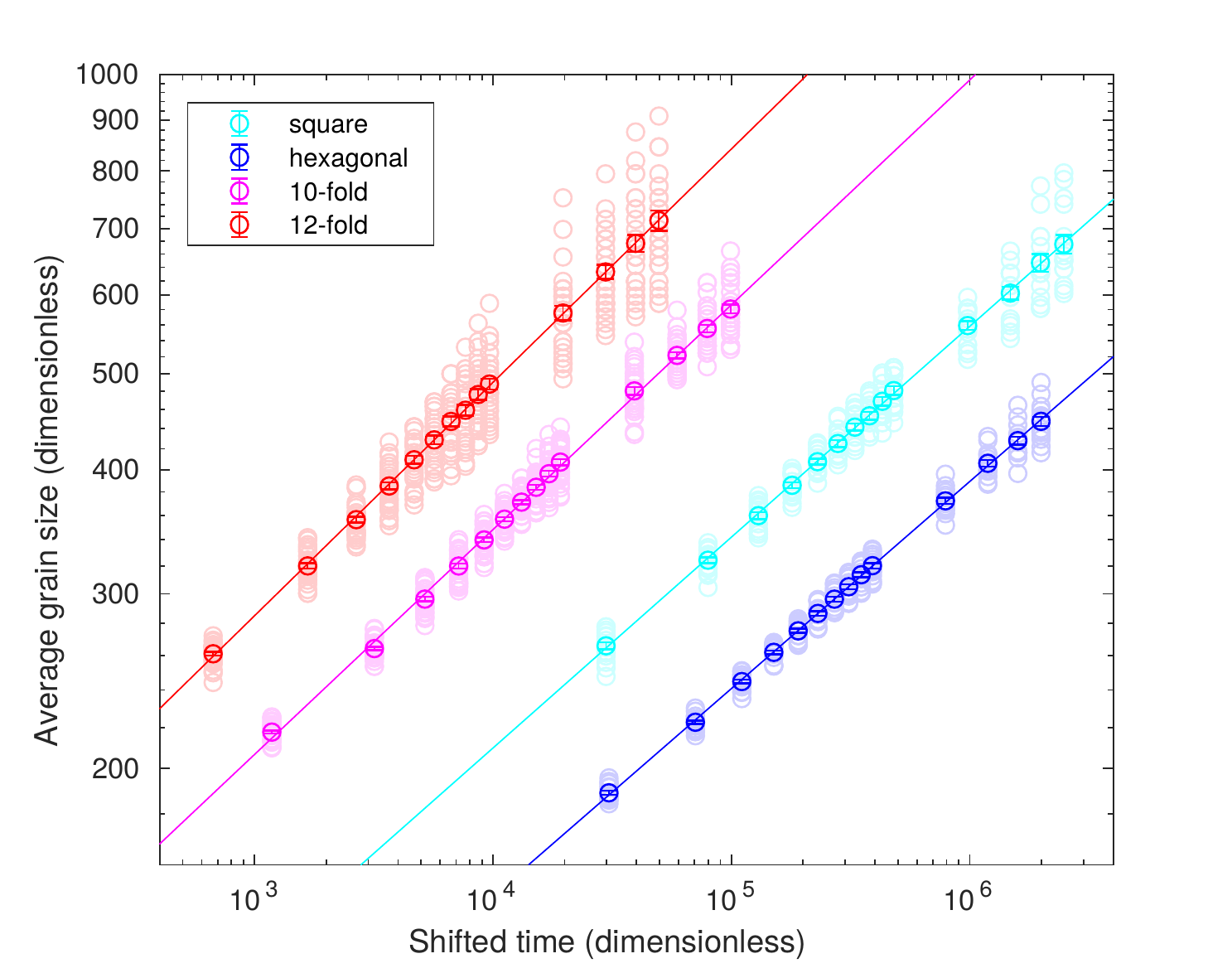}
\caption{Evolution of the average linear grain size averaged over the individual realizations as a function of shifted time for the four lattice types considered. See the main text for an explanation of the shifted time. The markers are actual data and the straight lines are power law fits. The fainter ghost markers show the average linear grain sizes of individual realizations.}
\label{fig-grain-size-evolution}
\end{figure}

As an archetypal benchmark of microstructural analysis, we first consider grain growth. Based on theoretical models \cite{ref-grain-growth-1949,ref-grain-growth-1952,ref-grain-growth-diffusion}, power-law growth is expected for the average grain size
\begin{equation}
\label{eq-grain-growth}
\left\langle d(t) \right\rangle = \alpha \left( t + t_0 \right)^\beta,
\end{equation}
where $\alpha, t_0$ and $\beta$ are fitting parameters, $\beta$ known as the growth exponent. While curvature \cite{ref-grain-growth-1949,ref-grain-growth-1952} and long-range diffusion \cite{ref-grain-growth-diffusion} driven growth correspond to well-defined universality classes of growth with $\beta = 1/2$ and $1/3$, respectively, PFC captures a more comprehensive picture of the microstructure, which incorporates numerous defect structures. We fitted our data of average grain sizes as a function of time with Eq. (\ref{eq-grain-growth}) to find $\beta$ for the different lattice types. Note that the relaxations were initialized with rather artificial tiled states, corresponding to different nonzero grain sizes at simulation time $t = 0$. Figure \ref{fig-grain-size-evolution} gives the evolution of the average grain size for the four lattice types as a function of shifted time where the time offset $t_0$ due to the nonzero initial grain size has been eliminated. Perfect power-law growth is observed for all lattice types with exponents $\beta = 0.21, 0.21, 0.23$ and 0.24. The hexagonal model used here is identical to that of Backofen \textit{et al.} \cite{ref-backofen}, and we obtain essentially the same growth exponent: our $\beta = 0.21$ \textit{vs.} their $\beta = 0.2$. Note that they originally reported $\beta_A = 2/5$ for grain area $A$ which corresponds to $\beta = 1/5$ for the linear grain size. The linear sizes of the model system are approx. 4500, 6600, 4100 and 3300 in dimensionless units, which are much larger than the corresponding average linear grain sizes even at $t / \Delta t = 5 \times 10^6$.

\subsubsection{Normalized grain size distributions}

\begin{figure*}
\centering
\includegraphics[width=\textwidth]{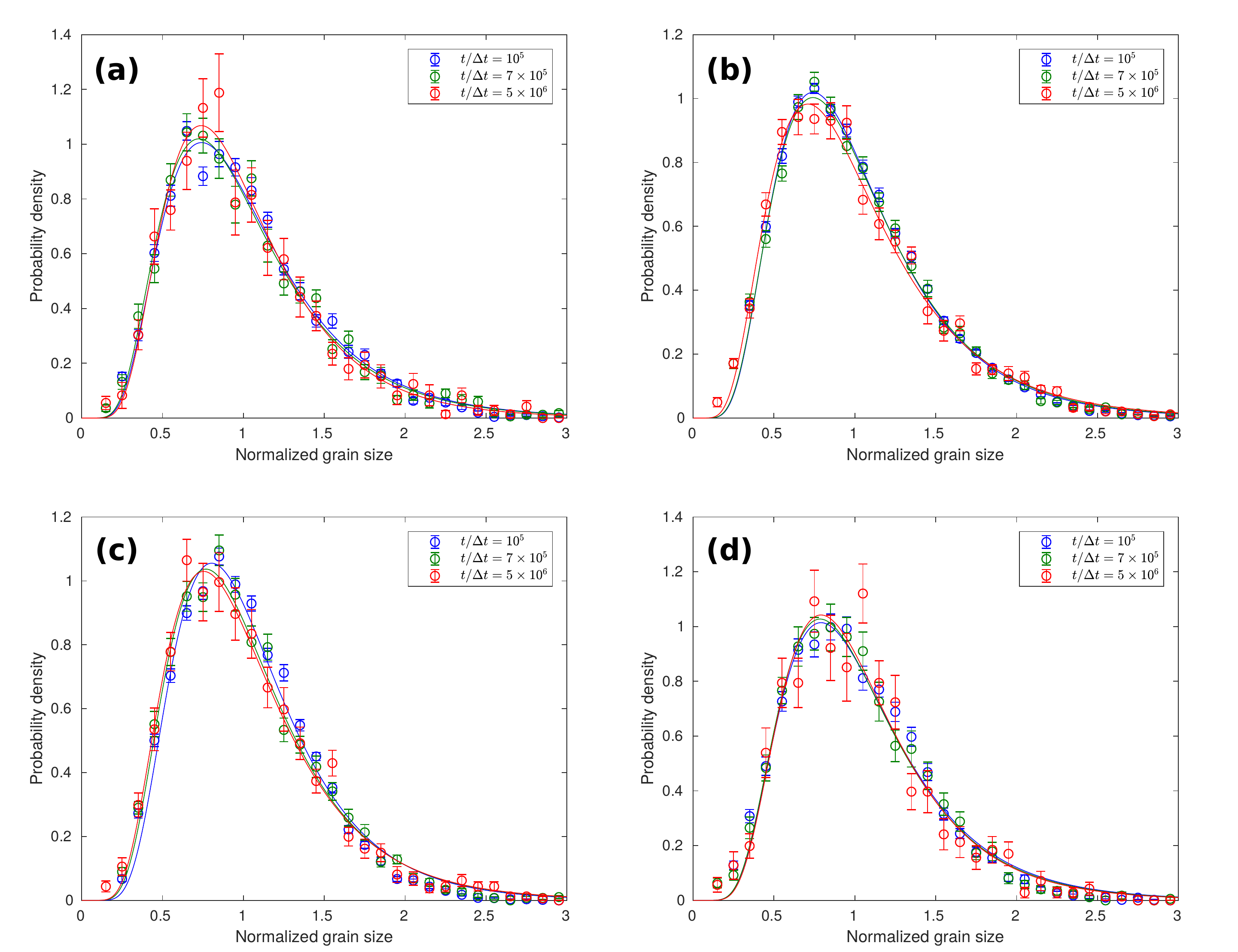}
\caption{Normalized grain size distributions for the four lattice types. Regular (a) square and (b) hexagonal, and (c) 10-fold and (d) 12-fold quasicrystal lattices. Three distributions are given at roughly exponentially spaced time steps. The markers are actual data and the curves are log-normal fits. The markers and error bars give the mean and the standard error, respectively, of the 16, 16, 32 and 32 parallel realizations of square, hexagonal, 10-fold and 12-fold model systems; recall the beginning of this subsection \ref{sec-analysis}.}
\label{fig-grain-size-dist}
\end{figure*}

Figure \ref{fig-grain-size-dist} shows the normalized grain size distributions $d/\left\langle d \right\rangle$, where the size of an individual grain $d = \sqrt{A}$, \textit{i.e.}, it is taken to be the square root of the grain's area $A$. The distributions appear log-normal as has been reported previously \cite{ref-barmak, ref-backofen, ref-gabriel}. A sufficient but not necessary cause for a log-normal distribution is a proportionate growth process \cite{ref-proportionate-growth}. However, it has recently been shown that a failure to detect low-angle grain boundaries can also result in detecting a log-normal grain size distribution where the true distribution is in fact different \cite{ref-pf-grain-size-dist}. While either or both may be the case here, the present grain extraction method was optimized to reproduce the segmentations determined by visual inspection by one of the authors (PH), wherein any error ultimately  lies with human judgment. On the other hand, the present data cannot confirm the observation that, for a hexagonal lattice, the distributions should become wider in time \cite{ref-gabriel}. There, an efficient numerical scheme \cite{ref-elsey-numerics} was used to push grain coarsening significantly further. Due to the greater computational workload, brought about by the four lattice types considered in this work, we limited ourselves to significantly shorter simulation times and can therefore neither confirm nor refute this observation. Regarding the different lattice types considered here, there are no obvious differences between them. The late time distributions display slightly more variance and these impaired statistics are due to larger, but fewer grains in the later systems. All the PFC distributions presented in this subsection are affected. Furthermore, the left-hand side tails are missing a couple of the leftmost data points in some cases, due to the size limit for extracting very small grains; recall Sec. \ref{sec-methods-grain-extraction}. All bins overlapping with the limit have been omitted.

\subsubsection{Grain size ratio distributions}

\begin{figure*}
\centering
\includegraphics[width=\textwidth]{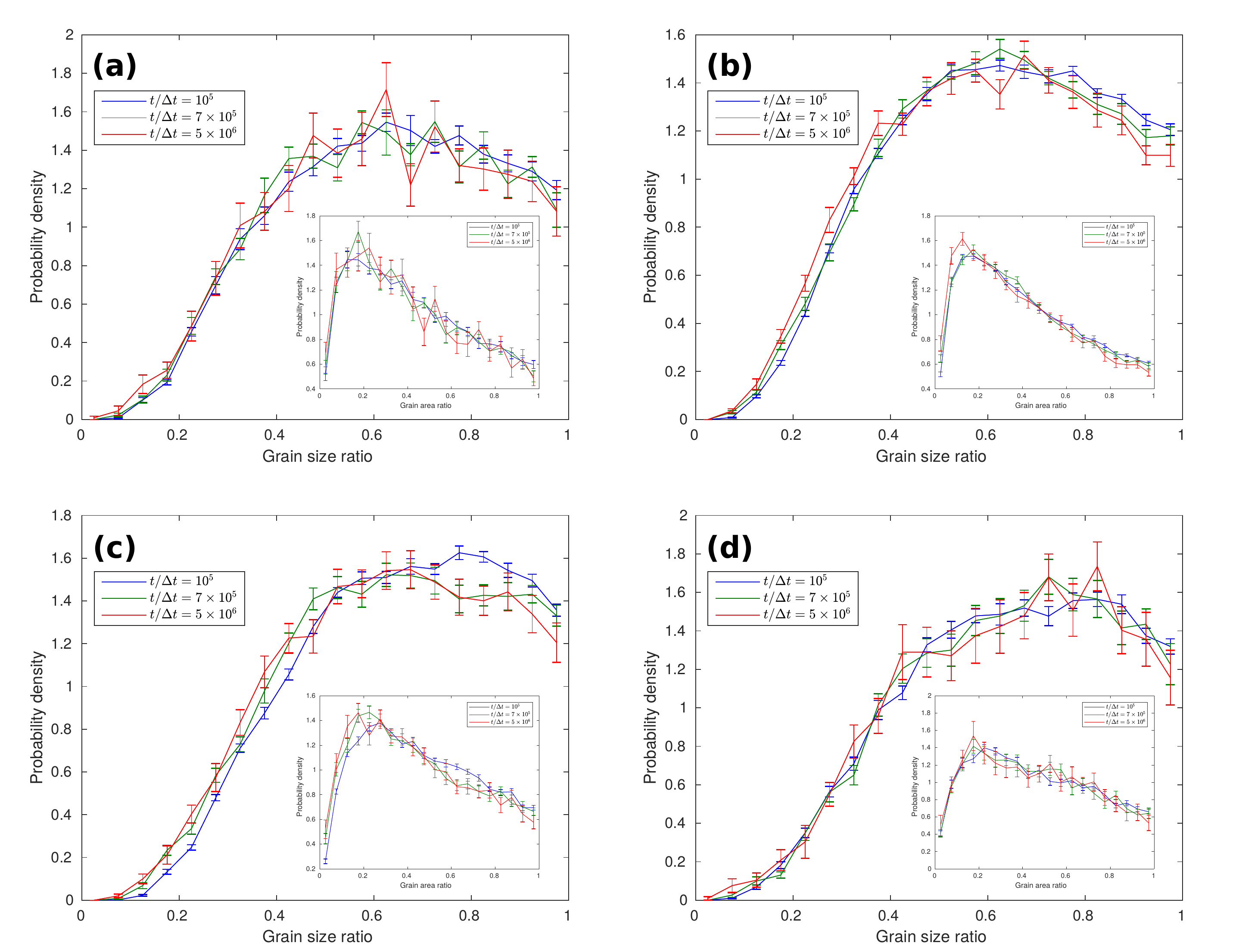}
\caption{Distributions of the linear size ratios of neighboring grains for the four lattice types. Regular (a) square and (b) hexagonal, and (c) 10-fold and (d) 12-fold quasi-lattices. Three distributions are given at roughly exponentially spaced time steps. The insets give the corresponding area ratios between neighboring grains. The lines are a guide for the eye.}
\label{fig-size-ratio-dist}
\end{figure*}

\begin{figure}
\centering
\includegraphics[width=0.5\textwidth]{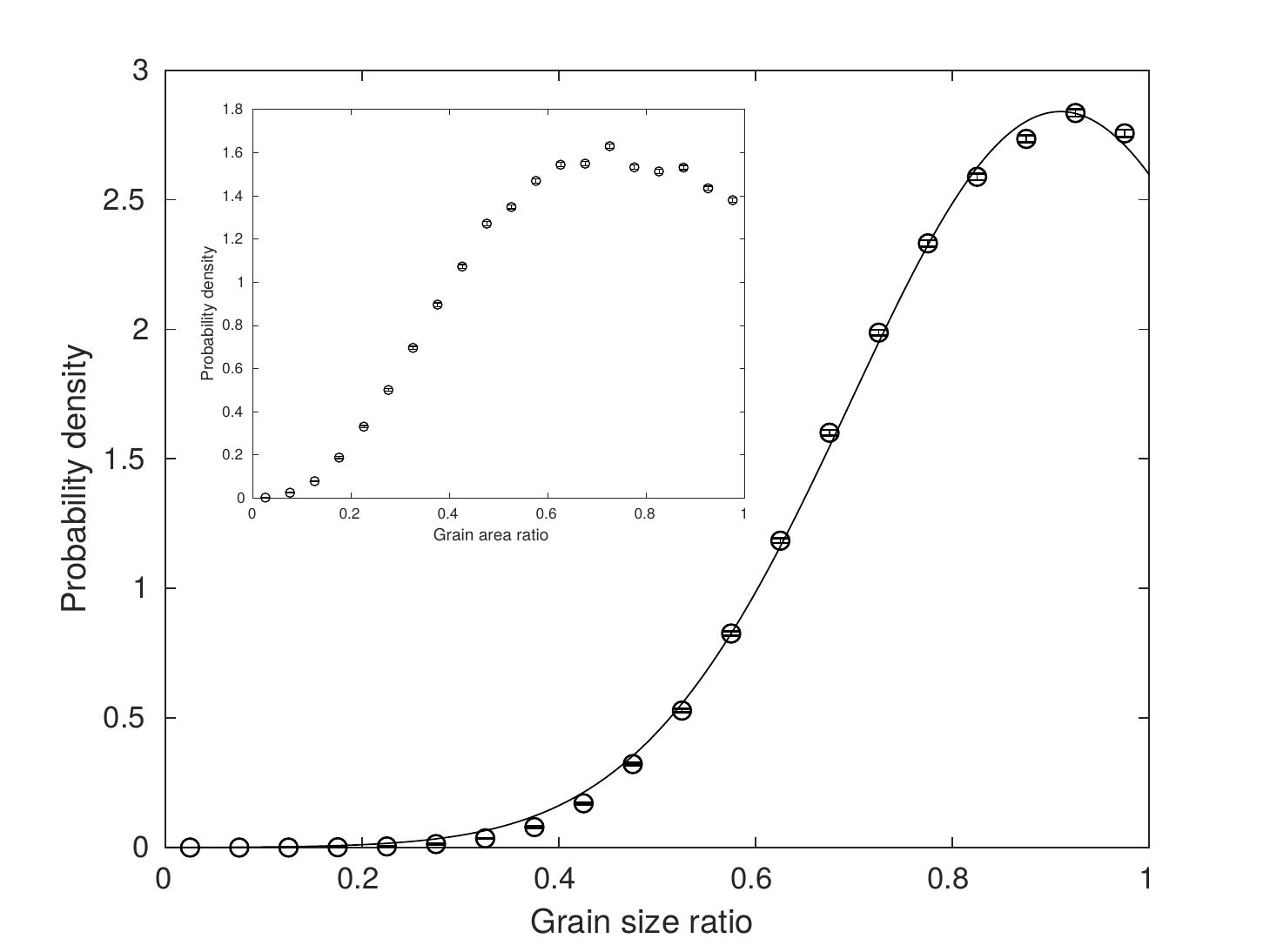}
\caption{Distribution of the linear size ratios of neighboring grains for Voronoi tessellations. The markers are actual data and the curve is an optimal truncated normal distribution fit. The inset gives the corresponding area ratios.}
\label{fig-size-ratio-dist-voronoi}
\end{figure}

Figure \ref{fig-size-ratio-dist} shows the linear grain size ratio distributions for the four lattice types. Linear grain size ratio is the ratio $\delta = d_{\min} / d_{\max}$, where $d_{\min}$ and $d_{\max}$ are the smaller and the larger, respectively, of the linear sizes of two neighboring grains. While moderate disparity in size seems preferred, the distributions are relatively flat for $0.4 \leq \delta \leq 1$. The average linear grain size ratios are 0.62, 0.61, 0.64 and 0.64 (at $t / \Delta t = 5 \times 10^6$). The insets give the corresponding distributions for the grain area ratio $\alpha = A_{\min} / A_{\max}$. All distributions appear similar and peak roughly around $\alpha \approx 0.2$, but the hexagonal distributions display slightly sharper peaks. The grain area ratio distributions for the hexagonal lattice are virtually identical to those reported in Ref. \cite{ref-gabriel}. The corresponding distributions for Voronoi grains in Fig. \ref{fig-size-ratio-dist-voronoi} are strikingly dissimilar. The distribution of the linear size ratios is quite close to a truncated normal distribution with an average of 0.80, and the area ratio distribution appears very different from the corresponding PFC distributions. It seems that random Voronoi tessellations give much less disparity in grain size, hinting of some essential physics related to grain growth dynamics that PFC is able to capture.


\subsubsection{Grain misorientation distributions}

\begin{figure*}
\centering
\includegraphics[width=\textwidth]{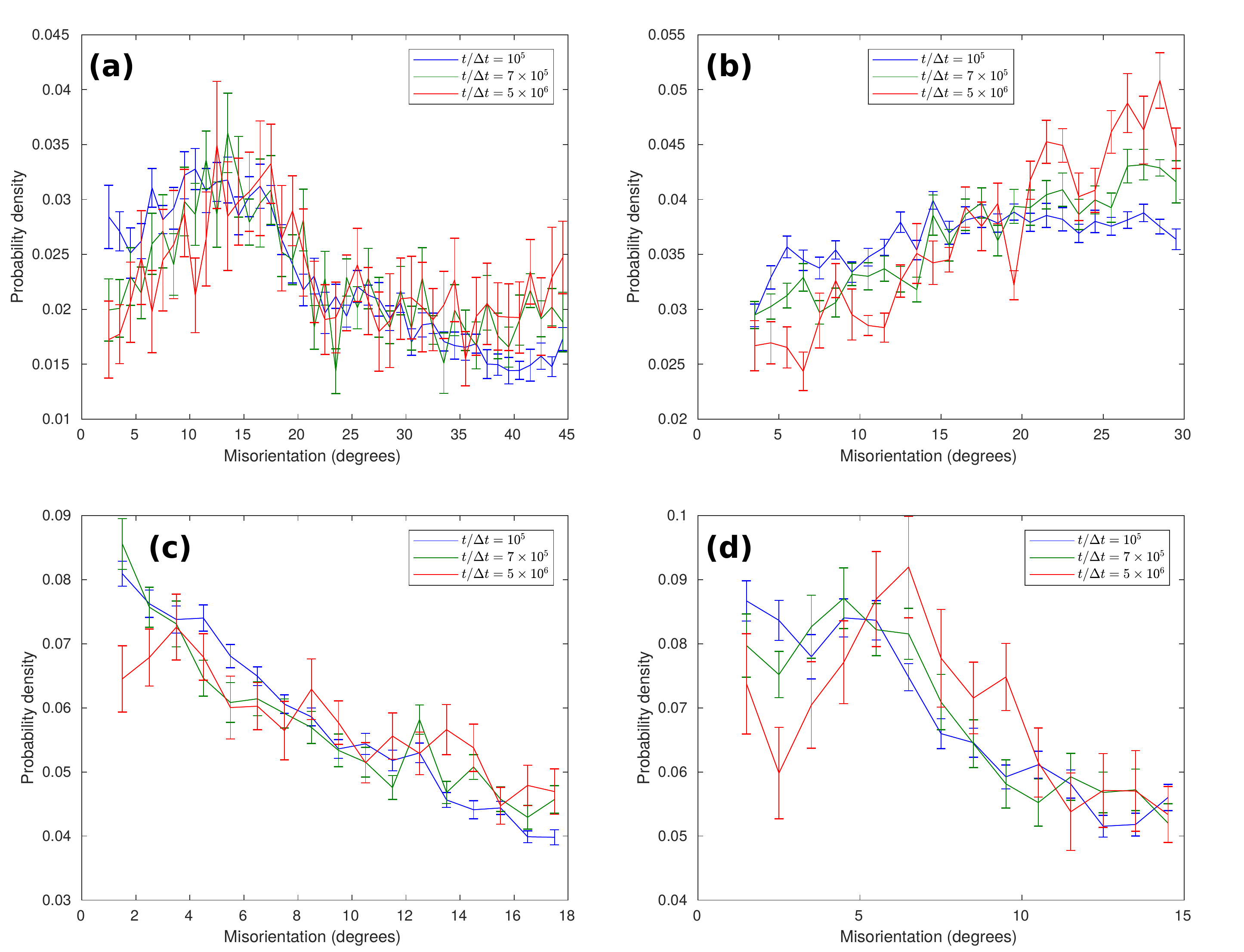}
\caption{Distributions of lattice misorientation between neighboring subdomains of neighboring grains for the four lattice types. Periodic (a) square and (b) hexagonal, and (c) 10-fold and (d) 12-fold quasicrystalline lattices. Three distributions are given at roughly exponentially spaced time steps.}
\label{fig-misorientation-dist}
\end{figure*}

Figure \ref{fig-misorientation-dist} shows the distributions of lattice misorientation between neighboring subdomains of neighboring grains for the four lattice types. Considering the misorientation between subdomains instead of grains (composed of, and their orientation averaged over, one or multiple subdomains) yields more accurate results. The frequencies of different misorientations have been normalized with corresponding grain boundary lengths. Note also that the maximal misorientations are $\theta = 45^\circ, 30^\circ, 18^\circ$ and $15^\circ$. All bins overlapping with the misorientation limits $\theta^\ast = 3.0, 2.5, 0.75$ and 0.5 have been omitted. The distributions appear very dissimilar between the four lattice types. The distributions for both the hexagonal lattice and the 10-fold quasi-lattice are approximately linear, but, surprisingly, the former gives more probability for larger and the latter for smaller misorientations. On the other hand, the distributions for the square lattice and the 12-fold quasi-lattice are not as trivial to characterize, but both display wide excess around $\theta \approx 15^\circ$ and $\theta \approx 7^\circ$, respectively.

Regarding hexagonal systems, a slight preference toward smaller misorientations has previously been reported \cite{ref-barmak, ref-gabriel}. The present method was used to analyze different time steps of a hexagonal model system used in Ref. \cite{ref-gabriel}. We confirm this conflicting preference toward smaller misorientations, whereby it appears that the misorientation distributions are dependent not only on the lattice type, but also on the model used and its parameters. This stands to reason, because the grain boundary energy --- a prime driver of microstructure evolution --- depends strongly on the PFC model \cite{ref-multiscale} and parameters \cite{ref-pfc-controlling-energy}. In addition, while Ref. \cite{ref-gabriel} reported an excess at $\theta \approx 10^\circ$, we do not see such a feature in our data, but, with extended simulation times, it could be possible for a corresponding bump to emerge. Last, qualitatively similar PFC models have been shown to predict energetically favored symmetrically tilted coincidence site lattice boundaries for misorientations $\theta \approx 18^\circ, 21^\circ$ and $28^\circ$ \cite{ref-multiscale}. The present data do display some excess for these misorientations, but, due to the relatively large error bars, we cannot conclusively distinguish these bumps from statistical fluctuations.

For the square lattice, we carried out grain boundary energy calculations using symmetrically tilted bicrystals to investigate the possibility of a connection between the features of the grain boundary energy and the misorientation distributions. However, the grain boundary energy measured appears very smooth and virtually featureless as a function of the tilt angle, and shows no kinks that could explain the excess observed at $\theta \approx 15^\circ$. We also investigated the grain boundary energies of symmetrically tilted grain boundaries for 12-fold quasicrystals, but again the energies obtained show no hints of particularly low-energy boundaries around $\theta \approx 7^\circ$. We must point out, however, that our analysis was not exhaustive and may have failed to detect hypothetical narrow kinks in grain boundary energy. In fact, unpublished results of author CVA show evidence of a possibly related kink at $\theta \approx 5.5^\circ$ which is in agreement with the interface dynamics when growing quasi-crystals from two seeds of different size \cite{ref-cristian-qc-interfaces}. Full details of the grain boundary energy calculations are given in Sec. \ref{sec-gb-calculations} in Appendix. More comprehensive investigation of quasicrystal grain boundary energies will be left for a future work. Before concluding on grain boundary energies, we would like to point out that the present grain extraction method does not distinguish between symmetric and asymmetric tilt boundaries of identical misorientation, and that the former are a special case of grain boundaries whereas the latter more general family of grain boundaries is much more abundant in the present microstructures. Unfortunately, investigating grain boundary energies with the additional degree of freedom brought about by asymmetric boundaries goes well beyond the scope of this work.

\subsubsection{Summary of additional results}

The rest of the microstructural results are given in full detail in Sec. \ref{sec-microstructure-further} of the Appendix. Table \ref{tab-results} lists all main results from this section. The grain aspect ratios, \textit{i.e.}, the ratio of the shorter principal axis to the longer, were found to be modest with averages 0.70, 0.66, 0.72 and 0.71 (at $t / \Delta t = 5 \times 10^6$), meaning that the most grains are slightly elongated. This is in reasonable agreement with random Voronoi tessellations with an average of 0.63. The aspect ratios are normally distributed. The grain misalignment, or the angle between the longer principal axes of two neighboring grains, shows tendency toward mutual alignment. In contrast, random Voronoi tessellations disfavor intermediate misalignments. We ascribe this difference to PFC's ability to capture the interactions and anisotropy of grain boundaries \cite{ref-multiscale, ref-tjs}. We observed reversed log-normal grain circularities
\begin{equation}
C = \frac{4 \pi A}{P^2},
\end{equation}
where $A$ is grain area and $P$ its perimeter, for all lattice types. The average circularities are 0.76, 0.75, 0.78 and 0.77 (at $t / \Delta t = 5 \times 10^6$), all slightly less circular than a square \cite{ref-circularity} due to grain elongation. All other lattice types except hexagonal show some finite size effects or vestiges of the artificial, tiled initial state as the distributions start off as not quite log-normal. It is surprising that, while all distributions for all other quantities at $t / \Delta t = 10^5$ have converged to their respective equilibrium shapes, the relaxation time scale for circularities can be longer. Distributions for the number of neighbors per grain are also log-normal with averages 5.99, 6.00, 5.99 and 5.96. More or less similar values have been reported for random Voronoi tessellations (6) \cite{ref-voronoi-neighbors}, PFC systems (6.0) \cite{ref-gabriel} and experimental systems (5.8) \cite{ref-barmak}.

\begin{table*}
\centering
\caption{Summary of results of the microstructural analysis of different lattice types. Results are reported for the periodic square and hexagonal lattices as well as for the 10- and 12-fold quasi-lattices. Results for random Voronoi tessellations are also given where applicable. Distribution types and corresponding average quantities are given where applicable. The averages are reported for $t / \Delta t = 5 \times 10^6$. The asterisks indicate that at least some of the corresponding distributions display finite size or time effects.}
\label{tab-results}
\begin{tabular}{lcc}
\hline
\hline
Average grain size & Growth exponent & \\
\hline
square & 0.21 & \\
hexagonal & 0.21 & \\
10-fold & 0.23 & \\
12-fold & 0.24 & \\
\hline
\hline
Normalized grain size distributions & Type &  \\
\hline
square & log-normal &  \\
hexagonal & log-normal &  \\
10-fold & log-normal &  \\
12-fold & log-normal &  \\
\hline
\hline
Grain size ratio distributions & Type &  Average \\
\hline
square & nontrivial & 0.62 \\
hexagonal & nontrivial & 0.61 \\
10-fold & nontrivial & 0.64 \\
12-fold & nontrivial & 0.64 \\
Voronoi & truncated normal & 0.80 \\
\hline
\hline
Grain misorientation distributions & Description & \\
\hline
square & excess around $\theta \approx 15^\circ$ & \\
hexagonal & linear, larger misorientations preferred & \\
10-fold & linear, smaller misorientations preferred & \\
12-fold & excess around $\theta \approx 7^\circ$ & \\
\hline
\hline
Grain aspect ratio distributions & Type & Average \\
\hline
square & truncated normal & 0.70 \\
hexagonal & truncated normal & 0.66 \\
10-fold & truncated normal & 0.72 \\
12-fold & truncated normal & 0.71 \\
Voronoi & truncated normal & 0.63 \\
\hline
\hline
Grain misalignment distributions & Description & \\
\hline
square & smaller misalignments preferred & \\
hexagonal & smaller misalignments preferred & \\
10-fold & smaller misalignments preferred & \\
12-fold & smaller misalignments preferred & \\
Voronoi & intermediate misalignments disfavored & \\
\hline
\hline
Grain circularity distributions & Type & Average \\
\hline
square & reversed log-normal$^\ast$ & 0.76 \\
hexagonal & reversed log-normal & 0.75 \\
10-fold & reversed log-normal$^\ast$ & 0.78 \\
12-fold & reversed log-normal$^\ast$ & 0.77 \\
\hline
\hline
Neighbor count distributions & Type & Average \\
\hline
square & log-normal & 5.99 \\
hexagonal & log-normal & 6.00 \\
10-fold & log-normal & 5.99 \\
12-fold & log-normal & 5.96 \\
\hline
\hline
\end{tabular}
\end{table*}

\section{Summary and conclusions}
\label{sec-conclusions}

In this paper we have introduced and comprehensively benchmarked an efficient and accurate method for extracting grains and analyzing the microstructure in 2D  poly and quasicrystalline solids. The present method was optimized for different periodic and quasi-lattices based on manual segmentations. A high level of agreement was achieved in all cases. We expect that the accuracy of the method could be further improved by utilizing machine learning techniques for the final subdomain merging step of the method. We also showed that the present method is applicable to molecular dynamics generated data of free-standing graphene. It should also be possible to modify the method to segment diffuse microstructures from phase field simulations. Generalizing this method to 3D lattices and quasi-lattices would be more complicated, but also extremely valuable.

We used the method to analyze the microstructures of various lattice types. We considered both regular periodic square and hexagonal lattices, as well as 10- and 12-fold symmetric quasicrystals. We studied the sizes, aspect ratios, circularities and neighbor counts of individual grains; also the size ratios, misorientations and misalignments between all pairs of neighboring grains. For the most part, we observed good agreement with previous works for the hexagonal lattice, and also very similar behavior between all four lattice types, suggesting that many microstructural properties are universal beyond lattice symmetry.

However, a particularly interesting case is that of lattice misorientation between neighboring grains. A previous work reported a slight preference toward smaller misorientations for hexagonal lattices, but we observed a preference toward larger misorientations. This issue was resolved by analyzing model systems used in the previous work -- we found the same preference toward smaller misorientations. This suggests that the distribution of misorientations is sensitive not only to the lattice type, but also to the exact model and its parameters being used. For square lattice and 12-fold quasicrystal, an excess of boundaries is observed with a misorientation of $\theta \approx 15^\circ$ and $\theta \approx 7^\circ$, respectively. We sought an explanation from grain boundary energy calculations and ruled out wide kinks in grain boundary energy as culprits of the excesses observed.

We expect the present work to be valuable in the study of both regular periodic crystals and quasicrystals. While phase field crystal has been used succesfully in the past to study quasicrystals, we have here demonstrated large-scale coarsening simulations of polyquasicrystalline microstructures. We have also presented a powerful new method for analyzing those microstructures. A very limited number of related approaches have been demonstrated for regular periodic crystals, but the present one is a first for quasicrystals, opening a new door for understanding quasicrystalline microstructures.

\appendix

\section{Appendix}
\label{sec-appendix}

\subsection{Further applications}
\label{sec-further-applications}

\begin{figure*}
\includegraphics[width=\textwidth]{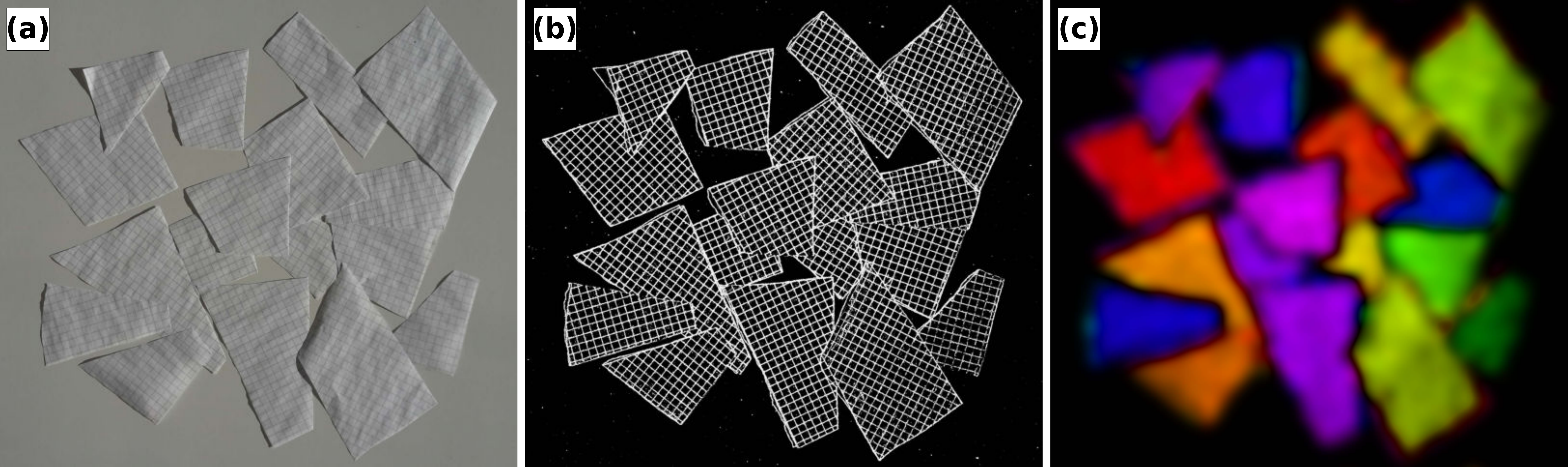}
\caption{Demonstration of the orientation field for scraps of cross-ruled paper. (a) The original photo, (b) a processed version used as the input and (c) the corresponding orientation field.}
\label{fig-paper}
\end{figure*}

\begin{figure*}
\includegraphics[width=\textwidth]{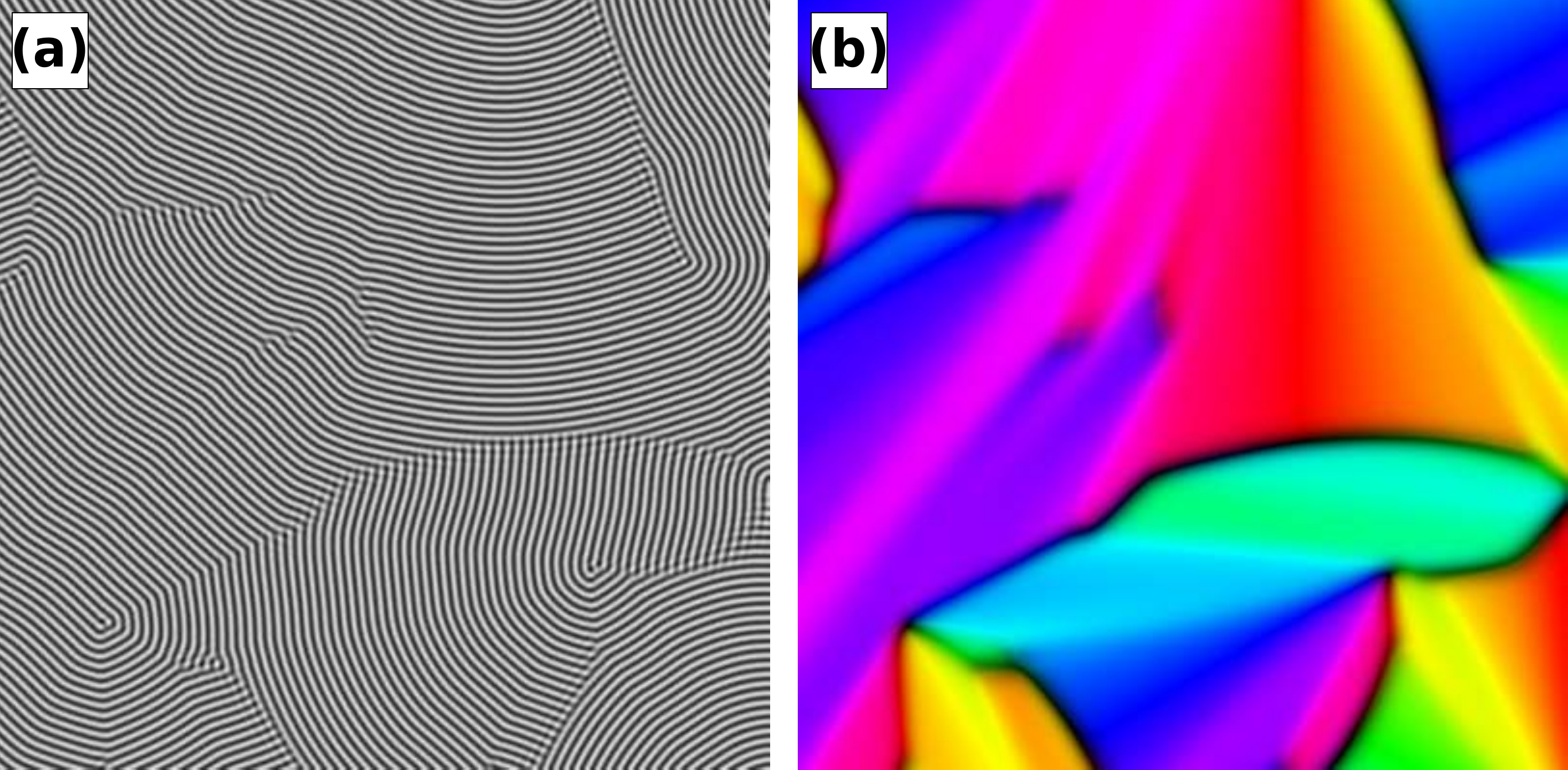}
\caption{Demonstration of the orientation field $\phi$ for a stripe system.}
\label{fig-stripe}
\end{figure*}

Virtually any 2D scalar field, or a norm of a complex-valued or a vector field, $\psi$ can be transformed into a corresponding orientation field $\phi$ assuming that $\psi$ displays periodic patterns with a fixed length scale and an even-fold rotational symmetry. As an example, Fig. \ref{fig-paper} demonstrates $\phi$ for scraps of cross-ruled paper. While the input data is somewhat unideal, the method does a good job of picking out the square grid orientations. A more realistic application could be to obtain the orientation field, or even the segmented grain networks, from experimental atomic-resolution images, such as from transmission electron \cite{ref-gb-mapping} or scanning probe microscopy. Optical images of colloidal systems could also be analyzed in this manner. As a further demonstration, Fig. \ref{fig-stripe} gives $\phi$ for a lattice symmetry class not considered in this work, a stripe system. Such structures are indeed relevant to colloidal systems \cite{ref-colloidal-stripe, ref-bacterial-stripe}, but also to surface ordering \cite{ref-fe-stripe}. Further potential fields of application could include image processing and machine vision.

\subsection{Computational performance of the grain extraction method}
\label{sec-performance}

The grain extraction method was implemented as two pieces of code, one being parallel C code and the other serial Java code. Computation of the orientation field $\phi$ and the deformation field $\chi$ are incorporated into the C code which utilizes MPI and FFTW for efficient parallel computation. The subdomain growth and merging steps, as well as analysis and visualization tools comprise the serial Java code.

We report here rough estimates of the execution times and memory usage for the largest system size considered in this work, namely $8192 \times 8192$ grid points. The square lattice system used for performance assessment had a total of 1084 grains. For benchmarking, the codes were run on a workstation with Intel Xeon E3-1230 v5 processor. Computation of $\phi$ and $\chi$ takes approx. 90 s (wall clock time) using the C code with 8 parallel MPI processes. Maximal memory usage was approx. 4.0 GB. With the serial Java code, the subdomain growth step takes approx. 90 s, whereas merging the grains only approx. 10 ms. Initialization and reading data take approx. 40 s. Tracing the grain boundaries and carrying out principal component analysis for the principal axes of the grains take approx. 3 s and 60 s, respectively. The maximal memory usage was approx. 10 GB.

We dedicate this paragraph to discussion on potential performance improvements. Due to the greater complexity of the subdomain growth and merging steps, they were implemented using the more user-friendly Java programming language. The serial subdomain growth step is the performance bottleneck -- the parallel computation of $\phi$ and $\chi$ can be sped up, up to some extent, by utilizing more CPU cores. Currently, the grid points are first sorted in the ascending order of $\chi$ using the sort method of the Collections Java class, and the grid points are then traversed in this order and assigned to the growing subdomains. A more efficient implementation would replace global sorting with local comparisons and would grow the subdomains on multiple fronts in parallel, starting from each local minimum in $\chi$. Another fairly obvious performance improvement would be to implement the method in a single piece of code. This would eliminate the current need to write and read $\phi$ and $\chi$ to and from disk. Lastly, coarse graining could be exploited to reduce both the execution times and memory usage. Significant downsampling of $\phi$ for computing $\chi$, and downsampling of both $\phi$ and $\chi$ for subdomain growth, could be feasible.

\subsection{Microstructural analysis of different lattice types: further results}

\label{sec-microstructure-further}

The grain extraction method was applied to study the microstructures of four different lattice types: square and hexagonal periodic lattices as well as 10- and 12-fold quasilattices. Major results are given in the main text, but we report here further related results. While the full details of these results are given here, they are also summarized in the main text. All values and error bars plotted here are the mean and the standard error, respectively, of parallel realizations of model systems.

\subsubsection{Grain aspect ratio distributions}

\begin{figure*}
\centering
\includegraphics[width=\textwidth]{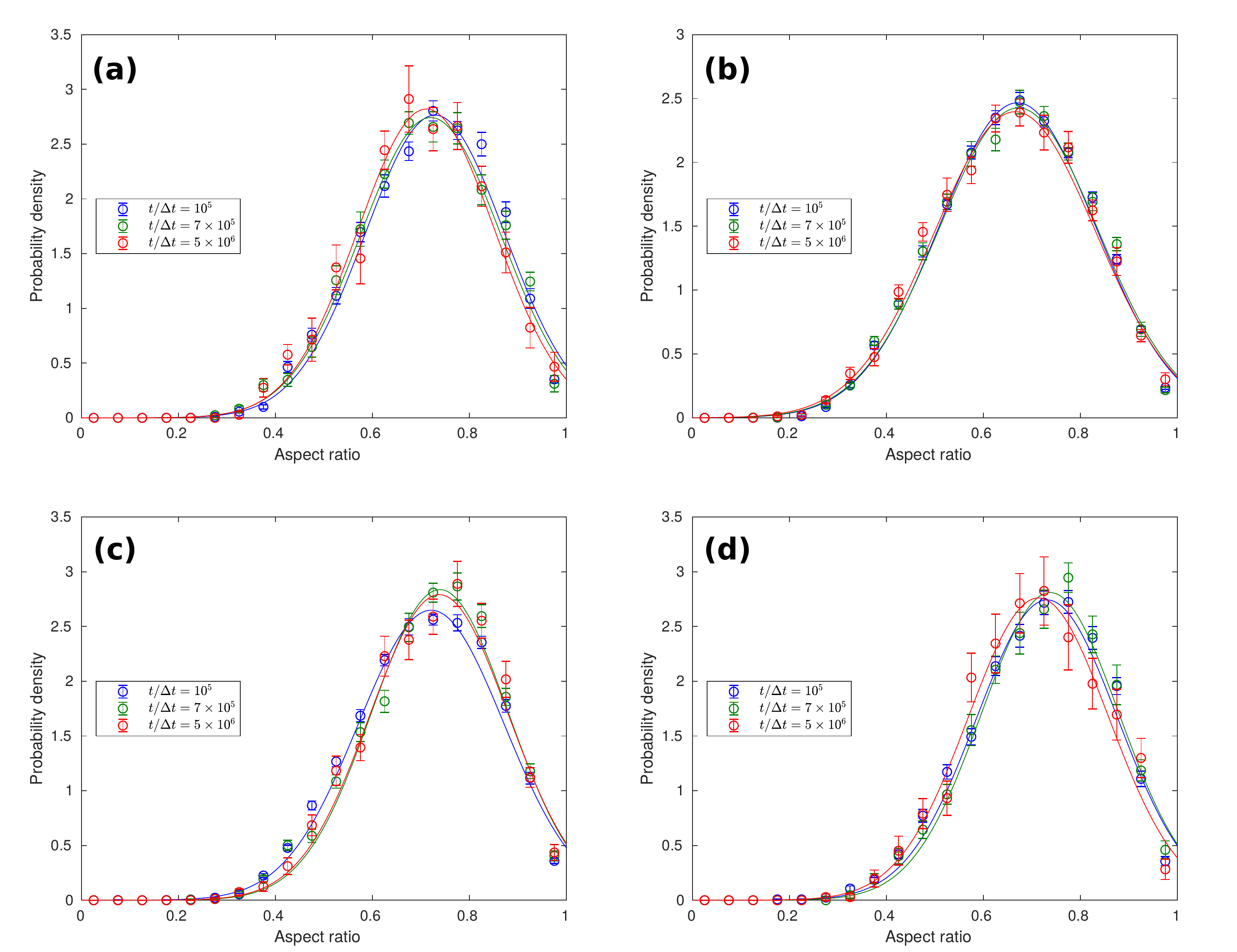}
\caption{Grain aspect ratio distributions for the four lattice types. Regular (a) square and (b) hexagonal, and (c) 10-fold and (d) 12-fold quasicrystalline lattices. Three distributions are given at roughly exponentially spaced time steps. The markers are actual data and the curves are optimal truncated normal distribution fits.}
\label{fig-aspect-ratio-dist}
\end{figure*}

\begin{figure}
\centering
\includegraphics[width=0.5\textwidth]{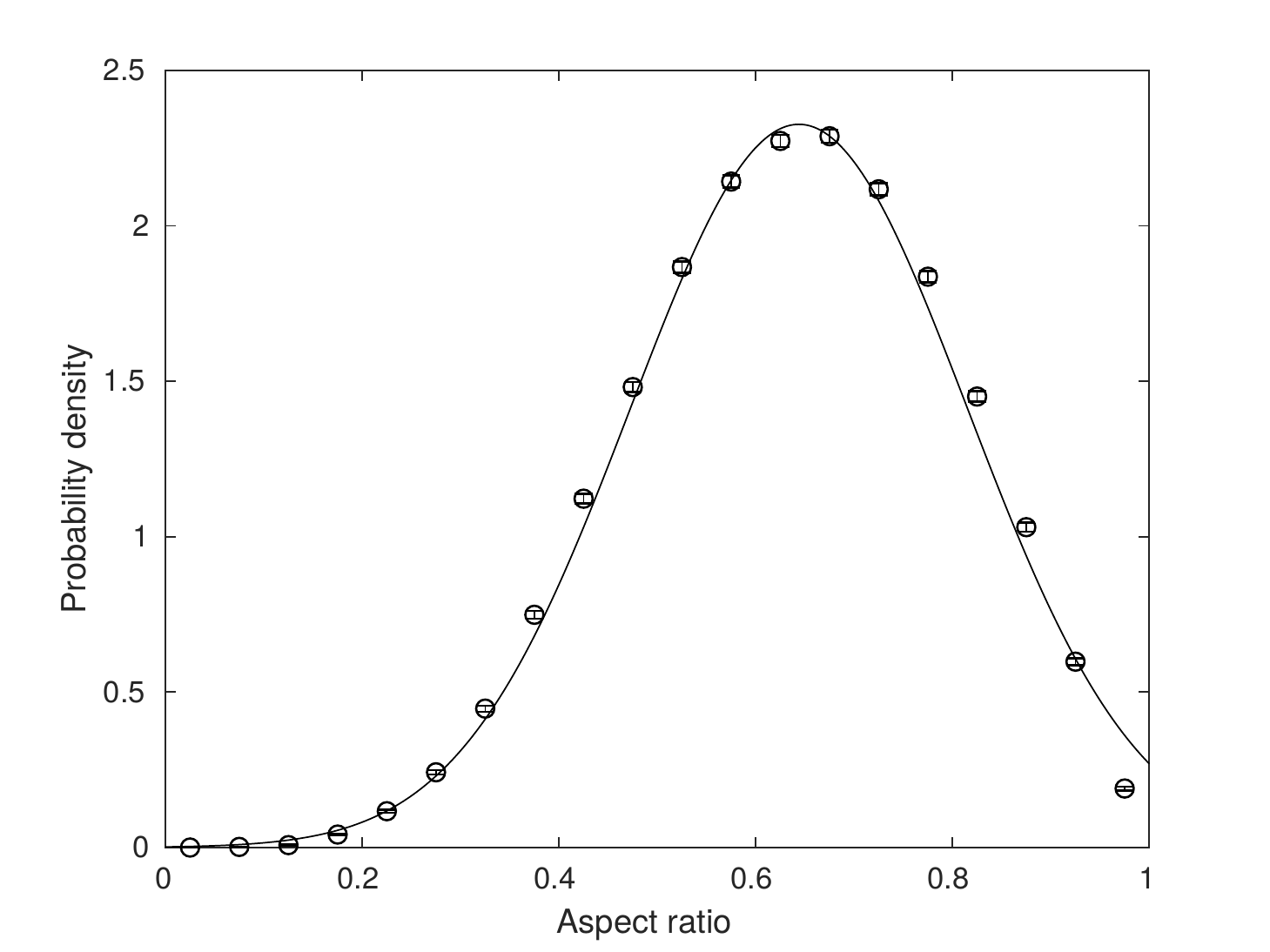}
\caption{Grain aspect ratio distribution for Voronoi tessellations. The markers are actual data and the curve is a truncated normal distribution fit.}
\label{fig-aspect-ratio-dist-voronoi}
\end{figure}

Figure \ref{fig-aspect-ratio-dist} gives the grain aspect ratio distributions for the four lattice types. Grain aspect ratio is the ratio of the lengths of its second (shorter) and first (longer) principal axes, given by principal component analysis. The length of a principal axis is the standard deviation in its direction of the grain's grid points from the grain's barycenter. While very low aspect ratios, or very elongated grains, are rare all lattice types favor some elongation of the grains with average aspect ratios of 0.70, 0.66, 0.72 and 0.71 (at $t / \Delta t = 5 \times 10^6$). The hexagonal grains are slightly more elongated compared to the other lattice types. The corresponding distribution for Voronoi grains in Fig. \ref{fig-aspect-ratio-dist-voronoi} appears very similar to the PFC ones with an average aspect ratio of 0.63. All distributions are approximated quite well by a truncated normal distribution, except for their right-hand side tails where the aspect ratios are capped to unity.

\subsubsection{Grain misalignment distributions}

\begin{figure*}
\centering
\includegraphics[width=\textwidth]{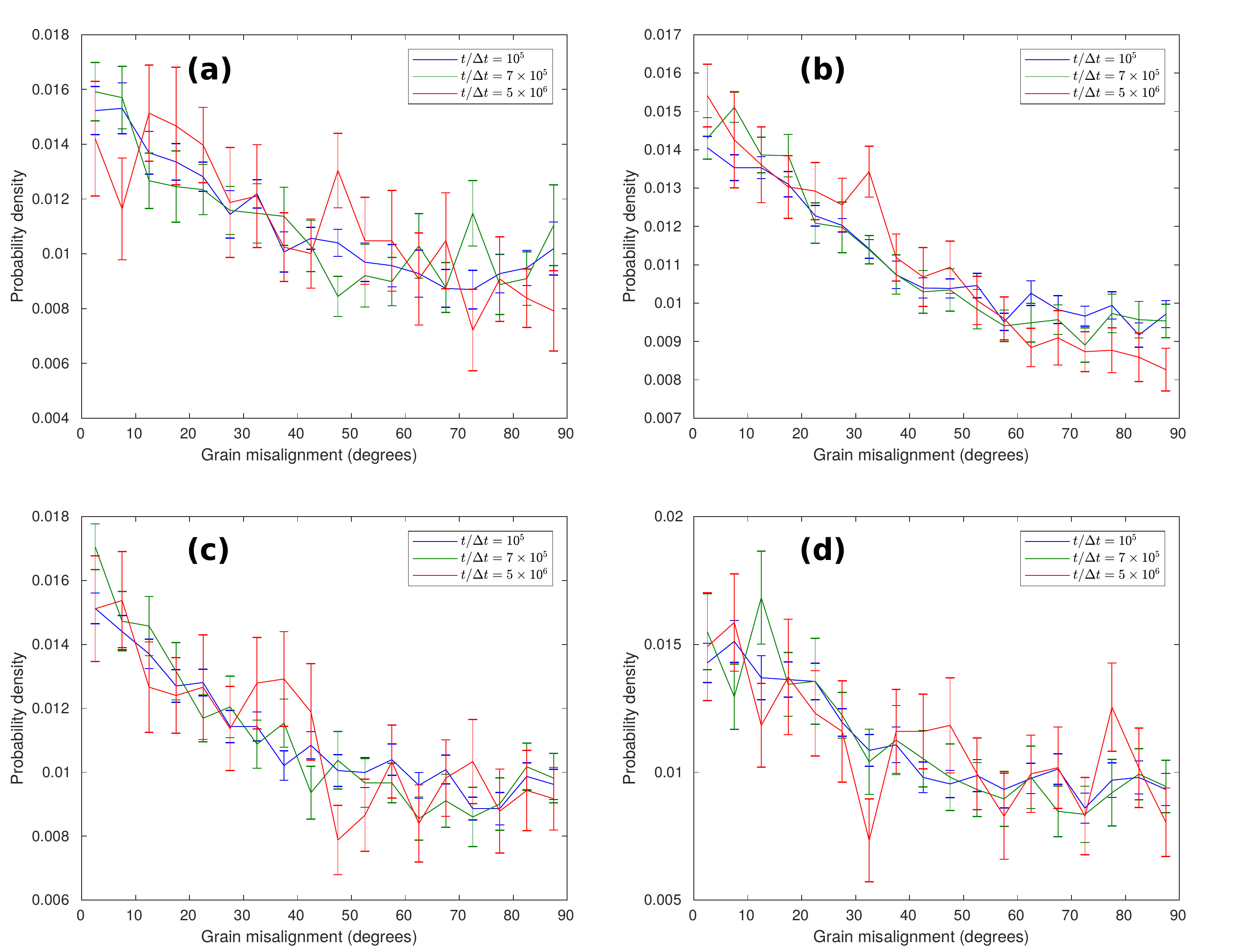}
\caption{Distributions of misalignment between neighboring grains for the four lattice types. Misalignment is the angle between the principal axes of two grains, \textit{i.e.}, the angle between the respective directions where they are the longest. Periodic (a) square and (b) hexagonal, and (c) 10-fold and (d) 12-fold quasicrystalline lattices. Three distributions are given at roughly exponentially spaced time steps. Only the pairs of neighboring grains where both have an aspect ratio $< 0.75$ have been included.}
\label{fig-misalignment-dist}
\end{figure*}

\begin{figure}
\centering
\includegraphics[width=0.5\textwidth]{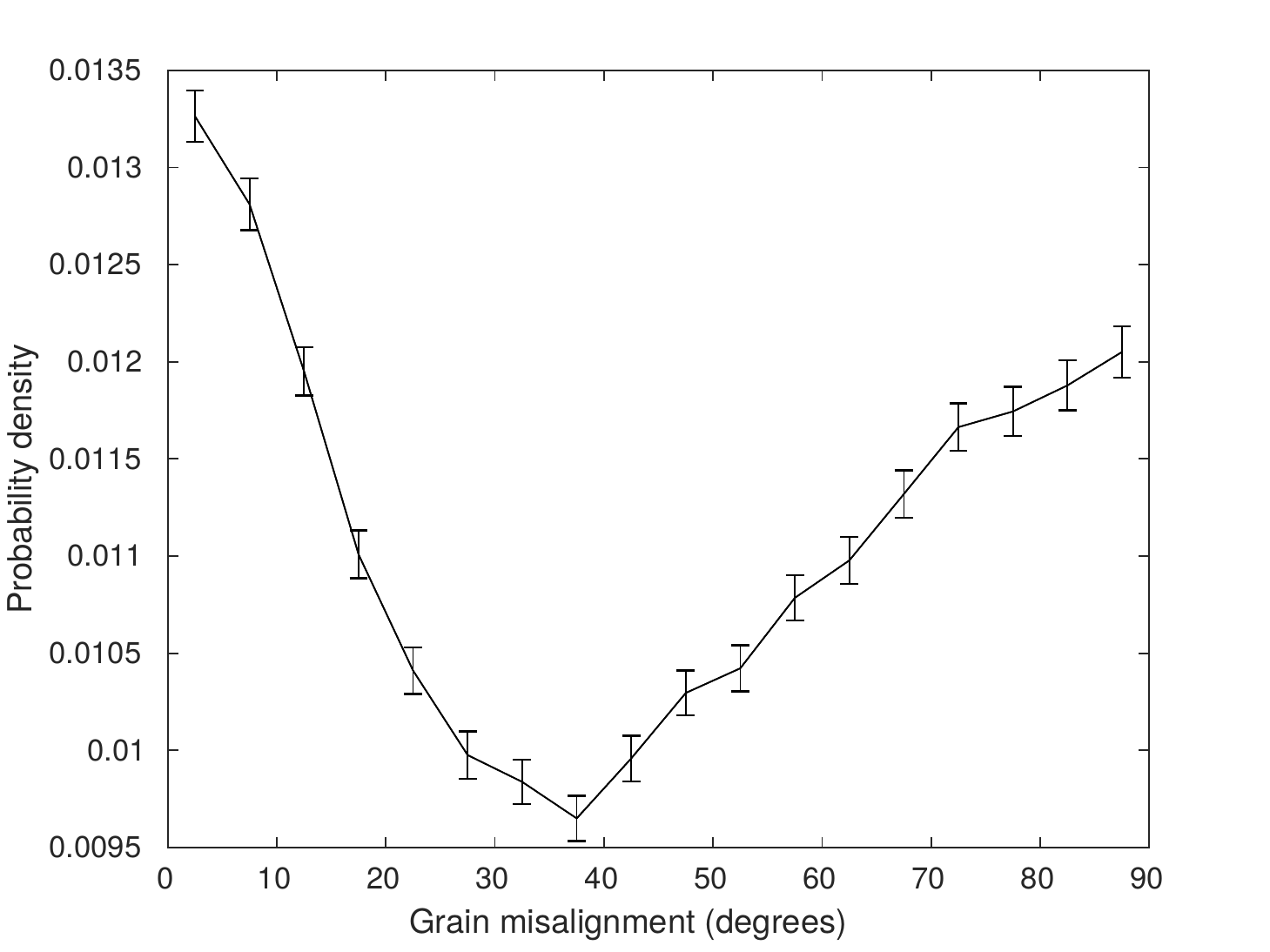}
\caption{Distribution of misalignment between neighboring grains modeled using random Voronoi tessellations. Only the pairs of neighboring grains where both have an aspect ratio $< 0.75$ have been included.}
\label{fig-misalignment-dist-voronoi}
\end{figure}

Figure \ref{fig-misalignment-dist} shows the distributions of misalignment between neighboring grains for the four lattice types. Misalignment between neighboring grains $\omega$ is the angle between their respective principal axes, \textit{i.e.}, the angle between the respective directions where the two grains are the longest. Note that $0^\circ \leq \omega \leq 90^\circ$. While most grains are at least somewhat elongated (see Fig. \ref{fig-aspect-ratio-dist}), only the pairs of neighboring grains where both have an aspect ratio $< 0.75$ have been included to limit the analysis to grains that are noticeably elongated. The distributions appear quite similar between all lattice types with some preference toward mutual alignment of neighboring grains. We investigated also the misalignment distributions of grains modeled using random Voronoi tessellations; see Fig. \ref{fig-misalignment-dist-voronoi}. The distribution for Voronoi grains is quite different with intermediate misalignments being disfavored. While simple random Voronoi tessellations correctly predict the log-normal distribution of grain sizes \cite{ref-voronoi}, it is clear that they do not give realistic grain misalignment distributions. Here there is actual physics involved, related to the interactions and anisotropy of the grain boundaries present, and PFC is able to capture such properties \cite{ref-multiscale, ref-tjs}.

\subsubsection{Grain circularity distributions}

\begin{figure*}
\centering
\includegraphics[width=\textwidth]{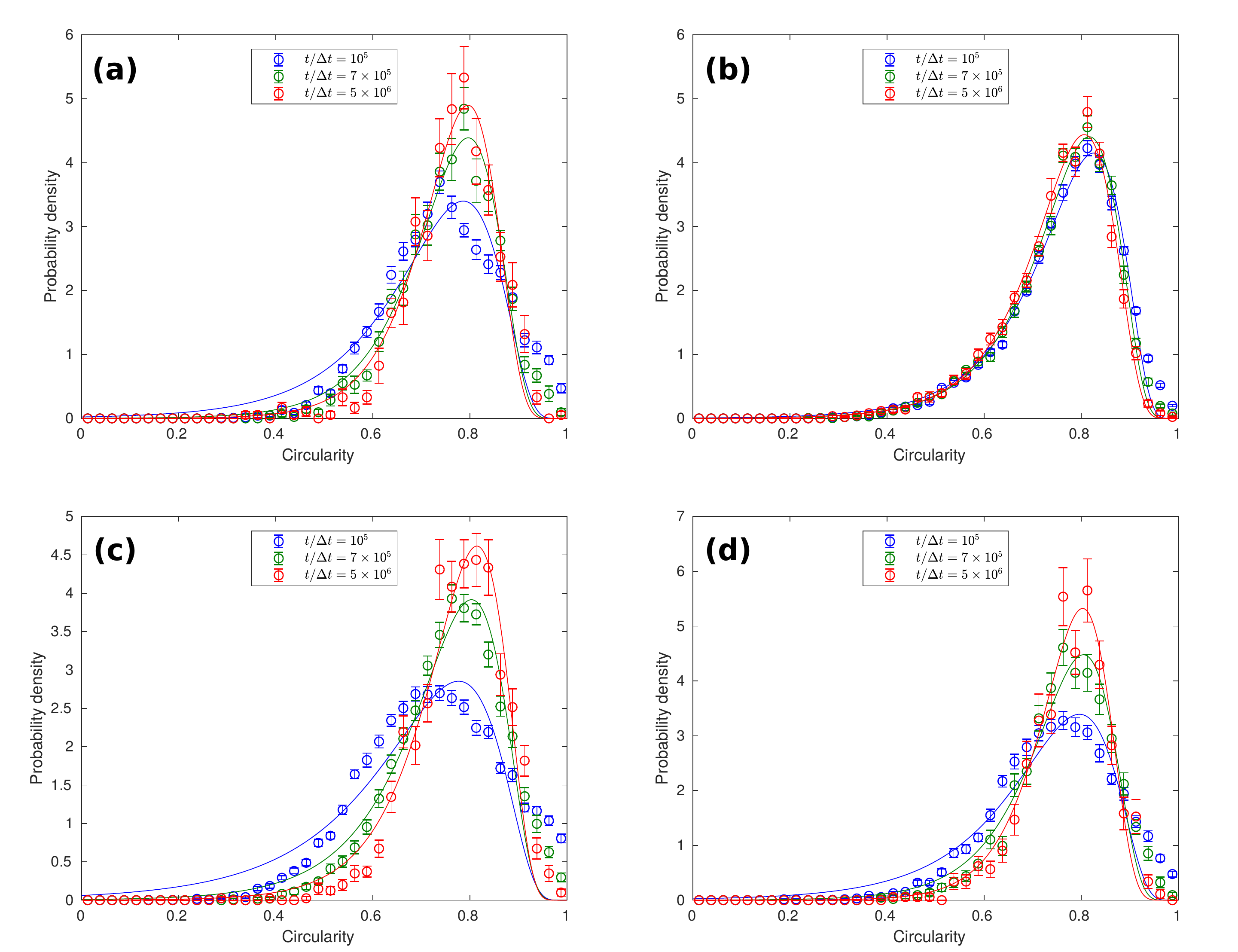}
\caption{Grain circularity distributions for the four lattice types. Regular (a) square and (b) hexagonal, and (c) 10-fold and (d) 12-fold quasicrystalline lattices. Three distributions are given at roughly exponentially spaced time steps. The markers are actual data and the curves are reversed log-normal fits.}
\label{fig-circularity-dist}
\end{figure*}

Figure \ref{fig-circularity-dist} shows the grain circularity distributions for the four lattice types. The grain circularity $C$ is given by
\begin{equation}
C = \frac{4 \pi A}{P^2},
\end{equation}
where $A$ is the grain area and $P$ its perimeter \cite{ref-gabriel}. A circularity $C = 1$ occurs only in circles while other shapes have $C < 1$. For example, a regular hexagon, a square and an equilateral triangle have circularities $C \approx$ 0.91, 0.79 and 0.60, respectively \cite{ref-circularity}. The perimeter or grain boundary length was measured by an algorithm that crawls along a boundary and uses a chain of line segments of length $a_\textrm{hex}$ to estimate its length. All late time distributions appear quite log-normal, just reversed, but the earlier time distributions for the square and quasicrystal lattices appear deviant. This is most likely due to finite size effects, or due to the quite artificial, tiled initial state. It is surprising that, while all distributions for other quantities at $t / \Delta t = 10^5$ have converged to their respective equilibrium shapes, the relaxation time scale for circularities can be longer. Because the algorithm can "cut corners", the circularities -- especially those of small grains -- are overestimated slightly. This effect is noticeable in the right-hand side tails of the earlier hexagonal distributions, but vanishes in the corresponding late time distributions. The average circularities are 0.76, 0.75, 0.78 and 0.77 (at $t / \Delta t = 5 \times 10^6$) -- all slightly less circular than a square. The grains are often somewhat elongated (see Fig. \ref{fig-aspect-ratio-dist}) which explains their relatively low average circularities.

\subsubsection{Neighbor count distributions}

\begin{figure*}
\centering
\includegraphics[width=\textwidth]{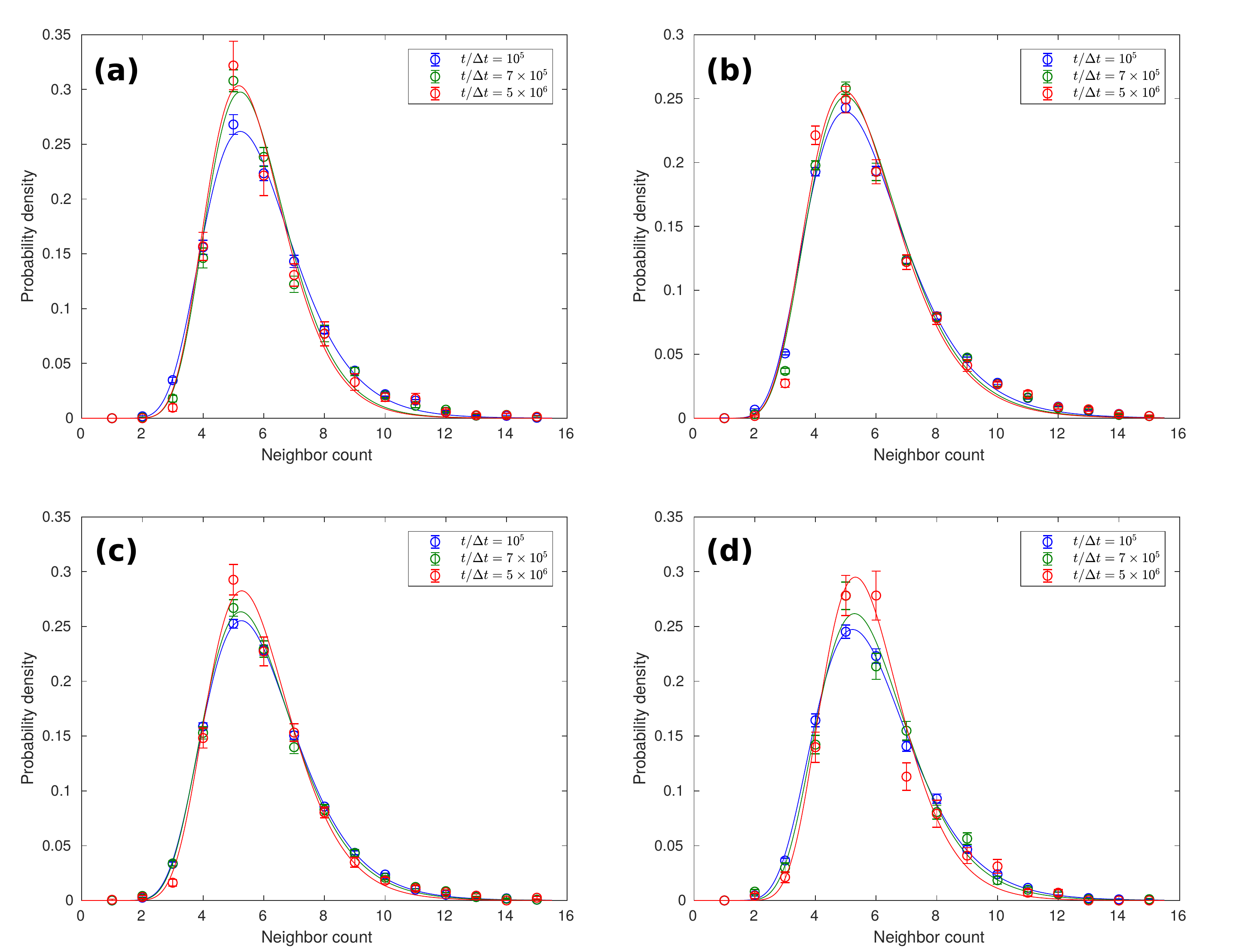}
\caption{Neighbor count distributions for the four lattice types. Regular (a) square and (b) hexagonal, and (c) 10-fold and (d) 12-fold quasi-lattices. Three distributions are given at roughly exponentially spaced time steps. The markers are actual data and the curves are log-normal fits.}
\label{fig-neighbors-dist}
\end{figure*}

Figure \ref{fig-neighbors-dist} shows the distributions of the number of neighbors of a grain for the four lattice types.

\subsection{Grain boundary energy calculations}

\label{sec-gb-calculations}

We calculated the grain boundary energy of symmetrically tilted boundaries for the square lattice to investigate if the excess of grain boundaries with misorientation $\theta \approx 15^\circ$ is caused by energetically favored boundaries at such misorientation. Similarly to our earlier work \cite{ref-multiscale}, we used bicrystalline model systems where the bicrystal halves were tilted symmetrically. The tilt angle is defined as $2\varphi = \varphi - \left( - \varphi \right)$, where $\varphi$ is the rotation angle of one grain from a reference orientation. Note that there are two tilt angles corresponding to every misorientation angle (except for maximal or zero misorientation) that correspond to different grain boundaries; \textit{cf.} armchair and zigzag grain boundaries in graphene \cite{ref-multiscale}. The grain boundary energy is given by
\begin{equation}
\gamma = \frac{L_\perp}{2} \left( f - f_\textrm{eq} \right),
\end{equation}
where $L_\perp$ is the length of the rectangle-shaped model system in the direction perpendicular to the grain boundaries, the factor $1/2$ comes from the fact that there are two grain boundaries in a periodic bicrystalline system, $f$ is the free energy density of the relaxed, defected bicrystal and $f_{eq}$ is the free energy density of the ground state. We considered a representative set of high-symmetry, or short period boundaries. The bicrystal halves were fitted with the computational unit cell in the direction of the grain boundaries. The bicrystal halves were initialized with a sharp interface. Possible finite size effects were assessed by doubling system dimensions; virtually identical results were obtained. The grain boundary energy appears very smooth and essentially featureless; no obvious downward kinks in energy, corresponding to low-energy boundaries that could explain the excess detected, are observed. Simulated annealing or perturbing the potentially metastable grain boundary configurations in other ways could possibly help reduce the energies, but we do not expect any pronounced kinks to appear in the energy.

\begin{figure}
\centering
\includegraphics[width=0.5\textwidth]{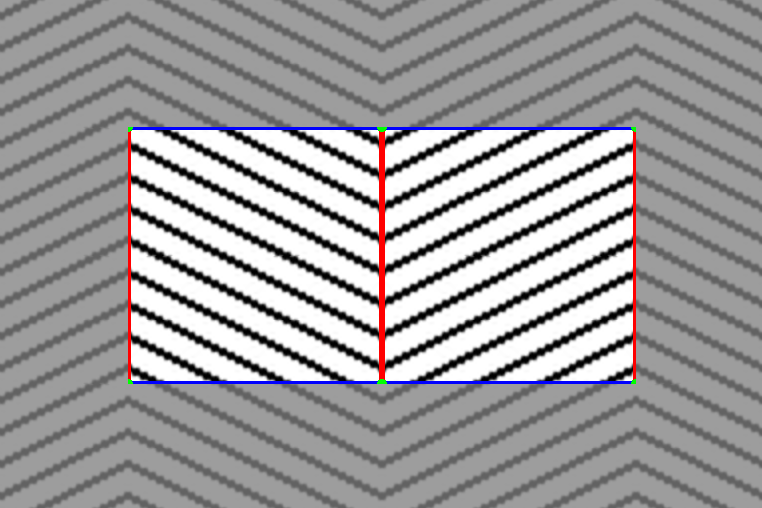}
\caption{Schematic of the biquasicrystal layout. The stripe pattern indicates an arbitrary, symmetrically tilted lattice orientation. The red lines indicate grain boundaries and blue lines phasonic boundaries. Green points correspond to quadruple junctions. The grayed-out margins show the continued periodic structure.}
\label{fig-bicrystal}
\end{figure}

A similar grain boundary energy analysis was carried out for the 12-fold quasicrystals to shed some light on the excess observed at $\theta \approx 7^\circ$. Calculating the grain boundary energies of quasicrystals in a similar manner is more complicated, because they cannot be fitted to a periodic calculation unit cell due to their aperiodic nature. As a consequence, phasonic flips will occur in the vicinity of the interfaces between a grain and its periodic images. A simplified schematic of the biquasicrystal layout is given in Fig. \ref{fig-bicrystal} where an arbitrary lattice direction is indicated by the stripe pattern. The total formation energy can now be expressed as
\begin{equation}
\label{eq-formation-energy}
F = f_\textrm{eq} S + 2 \gamma L_\parallel + \gamma^\ast L_\perp + 2 f_\textrm{p},
\end{equation}
where $S$ is the area of the system, $\gamma$ is the energy of the grain boundaries (highlighted in red in Fig. \ref{fig-bicrystal}), $L_\parallel$ is the length of the system in the direction of the grain boundaries, $\gamma^\ast$ is the energy of the phasonic flip boundary (blue) and $f_\textrm{p}$ is the formation energy of the quadruple junctions at the meeting points of the grain and the phasonic flip boundaries (green). To extract $\gamma$, we varied both system dimensions independently and the energy quantities $f_{\rm eq}$, $\gamma$, $\gamma^\ast$ and $f_\textrm{p}$ were solved by fitting Eq. (\ref{eq-formation-energy}) to the measured total formation energies using least squares. We scaled the computational unit cell dimensions independently by factors 1, 2, 3 and 4, and considered all permutations. We considered tilt angles $2\varphi = 1^\circ, 2^\circ, \cdots, 14^\circ$. Since the biquasicrystals could not be fitted with the periodic computational unit cell, it is expected that there is some rotation and strain due to mismatch, but these contributions should diminish as the system dimensions are scaled up. There is some uncertainty in the grain boundary energy as evidenced by the error bars of one-sigma confidence interval, but it is clear that the grain boundary energy does not display wide and deep kinks that could be related to the excess observed. It is possible, however, that hypothetical narrow kinks remain undetected between the tilt angles sampled --- rather sparsely due to the greater workload of the scaling analysis. A more comprehensive investigation of quasicrystal grain boundary energies is warranted.

\begin{figure*}
\centering
\includegraphics[width=\textwidth]{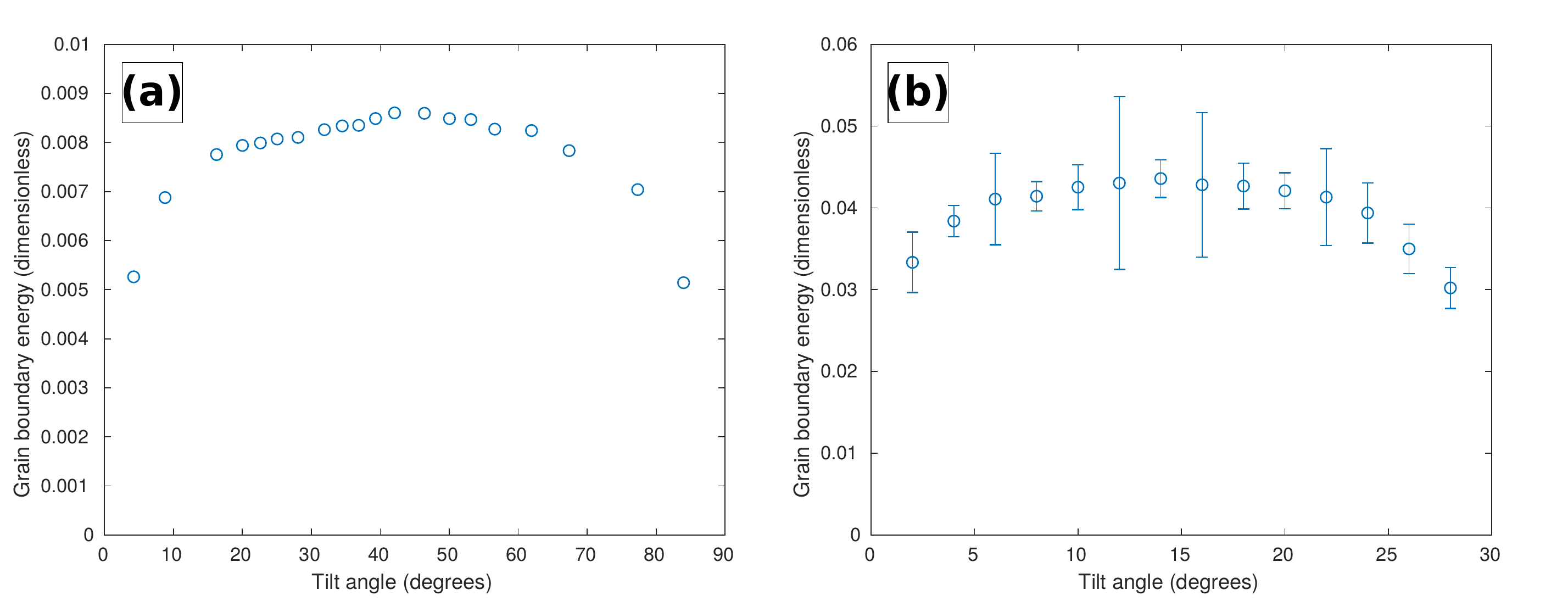}
\caption{Grain boundary energy of symmetrically tilted boundaries as a function of the tilt angle. (a) Square lattice and (b) 12-fold quasicrystal. The error bars in (b) correspond to one-sigma confidence intervals.}
\label{fig-gb-energy-4-12}
\end{figure*}

\bibliography{sample}

\section*{Acknowledgments}

This research has been supported in part by the 
Academy of Finland through its QTF Center of Excellence Program grant No. 312298.
We acknowledge the computational resources provided by the Aalto Science-IT project and the CSC IT Center for Science, Finland. PH acknowledges financial support from the Foundation for Aalto University Science and Technology, and from the Vilho, Yrjö and Kalle Väisälä Foundation of the Finnish Academy of Science and Letters. PH also wishes to thank Paul Jreidini and Matthew Frick for helpful discussions. GMLaB was supported by an FRQNT (Fonds de recherche du Qu\'ebec - Nature et technologies) Doctoral Student Scholarship. CVA acknowledges financial support from CRHIAM-FONDAP-CONICYT Project No. 15130015. NP  acknowledges financial support from the Canada Research Chairs (CRC) Program. K.R.E. acknowledges financial support from the National Science Foundation under Grant No. DMR-1506634 and from the Aalto Science Institute (ASCI).

\end{document}